Accessing the Free Expansion of a Crystalline Colloidal Drop by Optical Experiments.

Marcus Witt[1], G. H. Philipp Nguyen[2], Josefine R. von Puttkamer-Luerssen[1], Can H. Yilderim[1], Johannes A. B. Wagner[1], Ebrahim Malek[1,3], Sabrina Juretzka[1], Jorge L. Meyrelles Jr.[1,4], Maximilan Hofmann[1], Hartmut Löwen[2], and Thomas Palberg[1]

[1]Institute of Physics, Johannes Gutenberg University, Mainz, Germany

[2]Institute of Theoretical Physics II: Soft Matter, Heinrich-Heine-Universität Düsseldorf, Germany

[3]Institute for Biophysics, Goethe University Frankfurt, Germany

[4]Pontifícia Universidade Católica, Rio de Janeiro, Brazil

**Abstract**

We study poly-crystalline spherical drops of an aqueous suspension of highly charged colloidal spheres exposed to a colloid-free aqueous environment. Crystal contours were obtained from standard optical imaging. The crystal spheres first expand to nearly four times their initial volume before slowly shrinking due to dilution-induced melting. Exploiting coherent multiple-scattering by (110) Bragg reflecting crystals, time-dependent density profiles were recorded within the drop interior. These show a continuously flattening radial density gradient and a decreasing central density. Expansion curves and density profiles are qualitatively consistent with theoretical expectations based on dynamical density functional theory for the expansion of a spherical crystallite made of charged Brownian spheres. We anticipate that our study opens novel experimental access to density determination in turbid crystals.

## 1 Introduction

A block of solid composed of repulsive individual particles will immediately explode when exposed to a particle-free, vacuum-like environment. The reason are the strong interparticle forces or - in thermodynamic terms - the positive pressure of the solid, which needs to be counterbalanced to keep a solid stable. The resulting expansion process has been addressed, numerically and in experiments employing various model systems. In particular, charged plasmas and colloidal spheres are well suited for these investigations. Here, the repulsion type, range and strength are conveniently adjusted between the theoretical limits of hard spheres and the one component plasma [1, 2]. Moreover, the low shear rigidity of colloidal crystals allows their facile manipulation by external fields, like shear, gravity or gradients of chemical potential [3, 4, 5]. Starting with the pioneering work of Tang et al. [6], the majority of investigations studied expansion in two-dimensional systems. [7, 8, 9, 10, 11, 12, 13]. Many of these studies were motivated by and thus focussed on the 2D-specific two-step melting process induced by changes in density [14]. This is different in 3D, where expansion experiments have rarely been used to induce melting, which is a fascinating topic on its own [15, 16, 17] and can be conveniently induced by tuning the interaction strength [18, 19] or using temperature to induce particle size variations [20, 21].

In 2D, both density and refractive index mismatch between particles and suspending solvent allow for the facile preparation of experimental colloid systems, which are readily accessible by high resolution optical methods like confocal microscopy or light scattering [22, 23]. Moreover, for small two-dimensional samples, solvent currents arising during the initial placement or release process are typically too weak to affect the crystal evolution. By contrast, 3D expansion experiments in general require to start at elevated densities, where optical access is hampered by multiple scattering effects and sample turbidity. Further, isotropic expansion may be difficult to realize. The few reports available on the expansion of 3D crystals concentrated on unidirectional expansion after unidirectional compression [24]. Kanai et al. [25, 26] studied the expansion of centrifuged charged sphere crystals. They reported an approximately linearly decreasing density profile with discontinuity at the melting transition at values closely matching the equilibrium melting and freezing densities.

Isotropic expansion in three dimensions so far has been studied only for small plasma systems, which are still optically accessible [27]. Here, electrostatically driven collective effects dominate the expansion. Plasma systems come in two main variants, differing by repulsion type. Dusty plasmas in µg environments or thermophoretic levitation can form crystalline Yukawa balls (YBs). Laser-cooled ion plasmas released from harmonic trap potentials form expanding Coulomb balls (CBs). The density profiles of CBs are strictly flat at continuously decreasing densities and feature sharp boundaries, while those of YBs peak in the centre and have smeared boundaries [28, 29]. Presumably due to the challenges involved preparing well-defined starting conditions and the difficulties in optical access, experiments on the isotropic 3D expansion of colloidal solids have not been reported yet and experimental density profiles are missing.

The present study aims at closing this gap. We conduct expansion experiments on crystalline drops of highly concentrated crystalline colloidal suspensions in contact with a particle-free environment. The drops initially expand and later shrink by melting outward-in. They are overall turbid but show a brilliant colour display, which leaves their outer region and their contour accessible to various optical methods. The large central density, however, inhibits direct observation of the drop interior. We therefore developed an imaging method drawing contrast from the pronounced wavelength-dependent multiple scattering occurring within thin concentric shells of constant density, and matching the Bragg condition for (110) scattering. The approach allowed to deduce a considerable section of the radial density profiles as a function of time and yields density-dependent expansion curves.

To check the consistency of our heuristic concepts and to compare our data to theoretical expectations, we model the expansion by calculations based on dynamical density functional theory [30]. There, we consider particles interacting by DLVO-like Yukawa pair potentials. As starting configuration, we assume an isotropic expanding, spherical crystallite of body centred cubic crystal structure as well as a large and homogeneous initial density. Our results show a generally good qualitative agreement of the experimental data and this theoretical model. Few remaining discrepancies indicate room for further experimental and theoretical improvement. Overall, we successfully demonstrate a novel approach to time-dependent density profiling in expanding turbid media. We anticipate that multiple scattering based density determination will be useful in many future investigations on concentrated crystalline colloidal materials.

In what follows, we first introduce the materials and experimental techniques employed for drop fabrication and optical analysis. The following results section presents the experimental data. A short theoretical chapter presents our simple model and compares its expectations to the experimental data. In the discussion section, we address open points and remaining challenges. We conclude with some prospects for future applications of the demonstrated approach for improved expansion experiments.

## 2 Materials and experimental methods

### 2.1 Sample and sample characterization

Crystalline drops were prepared from highly charged polymer spheres in aqueous suspension. Our particles were a kind gift from BASF, Ludwigshafen (Batch no. GK2876 4542, Lab code PnBAPS80). From SAXS-measurements, their diameter and their polydispersity-index are $d$ = 87.5 nm and PI = 0.08, respectively. Conductivity measurements [31] yielded a conductivity charge of $Z_\sigma$ = 513±3, Torsional Resonance Spectroscopy gave an interaction effective charge of $Z_G$ = 365.1±2.3 [32, 33] (see also Fig. S1 in the SI). The particle mass density was $\rho_{mass}$ = 1.005 g/cm$^{-3}$. Deionized and decarbonized suspensions were prepared using standard procedures [34, 35]. In this state the systems are fluid at low and crystalline at large number densities, $\rho$, respectively. From reflection spectroscopy [36], their crystal structure is body centred cubic (bcc) up to the largest densities probed (Fig. S2 in the SI). From crystal growth measurements [37], the melting density is $\rho_m$ = (1 ± 0.2) µm$^{-3}$ under deionized, decarbonized conditions and $\rho_m \approx$ 15 µm$^{-3}$, after equilibration against ambient air [38]. Model calculations on deionized and decarbonized systems show little variation of the interaction strength over the density-range of interest (Fig. S3 in the SI).

### 2.2 Drop fabrication

For the expansion experiments, we prepared a stock of deionized and decarbonized suspension at $\rho$ = 110 µm$^{-3}$ >> $\rho_M$. A small amount of it was filled with some ion exchange resin [AmberjetTM Nr. K306.2, Carl Roth, Germany] into a syringe featuring an extrusion needle of inner diameter of 1.0 mm [Needle blunt 1.2×40mm, Braun, Germany]. After extruding a sacrifice-drop (not shown), we place the syringe tip some mm above the surface of deionized water contained in a four-side-polished optical cell with a square cross-section of 10x10 mm$^2$. Then we extrude another drop into the water and start the observation. A scheme of our experimental set-up is shown in Fig. 1a.

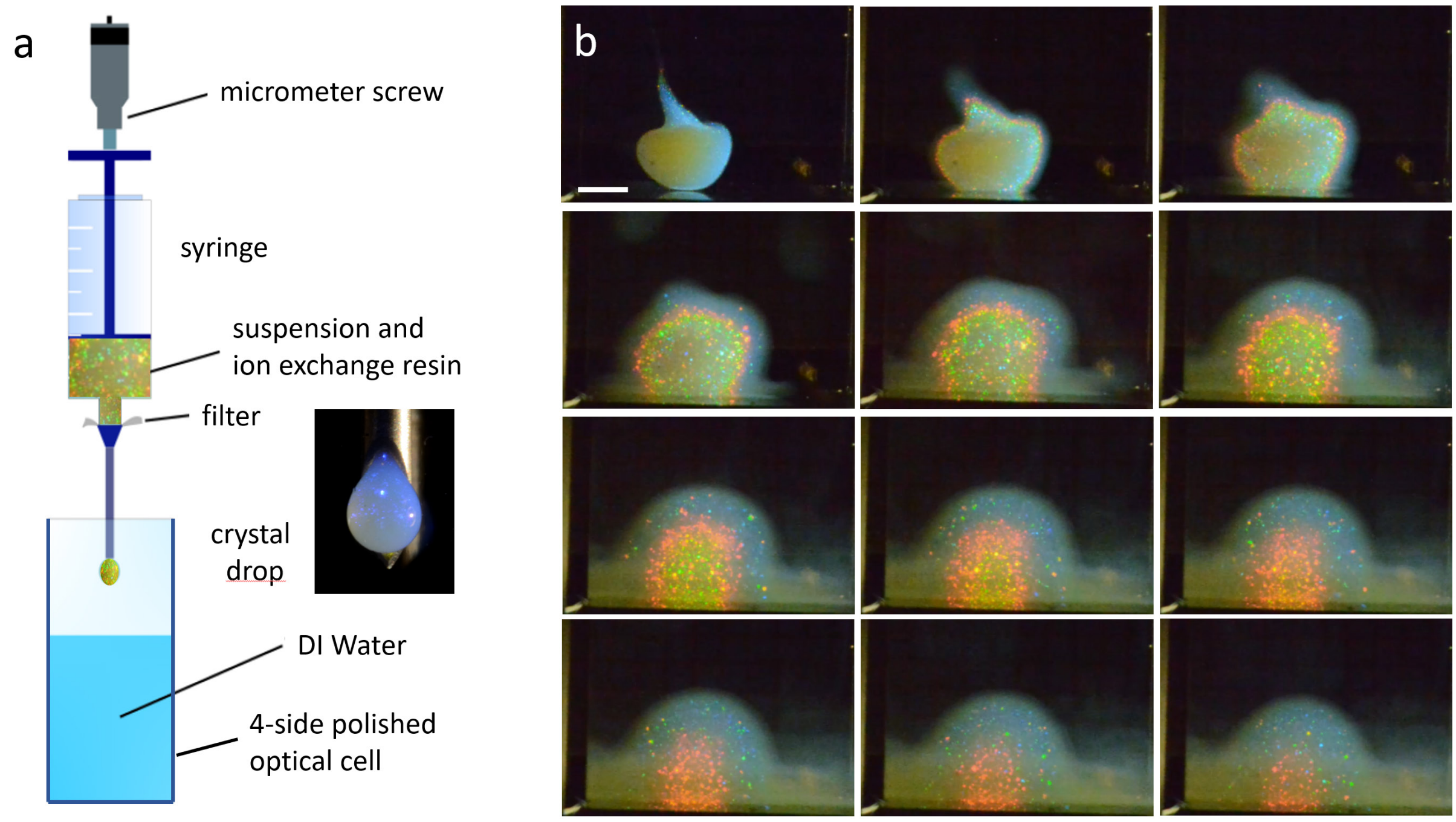

**Fig. 1 Setup and example Bragg images**. a) Arrangement of the dropping facility in side view. The inset shows a crystallized drop of suspension extruding from a syringe with an oblique tip. b) Time series of unprocessed Bragg images taken under white light illumination from the right. The time lag between images is $\Delta t \approx 60$ s, the scale bar is 2 mm. Note i) the characteristic colour pattern of the overall turbid drop, ii) the initial pointed tip at the upper drop side and the subsequent transition from anisometric to near spherical shape, and iii) the initial expansion, which is later followed by a slow shrinking of the crystalline region due to inward melting.

Due to the low shear modulus, the drop is shear-molten during the extrusion process, but it starts to recrystallize immediately after leaving the needle tip, i.e., within $t \approx 0.3$ s. Shape stabilization, however, take somewhat longer. For spherical drops, this occurs on average around $t \approx 5$ s. Since the initial placement history strongly influences the drop shape, it may take some manual skill and practice to obtain compact, spherical drops (cf. Fig. S5 of the SI). Due to the slight density mismatch with the environment, drops settle to the cell bottom. They initially expand and later shrink due to outward-in melting. Throughout, they show a brilliant colour display. The characteristic sequence of visual appearance is depicted in the unprocessed Bragg images shown in Fig. 1b (see also Video 1. For processed images, see below Fig. 3 and Fig. S9 in the SI).

### 2.3 Optical properties of drops and observation schemes

Our approach to density profiling in turbid systems relies on both light scattering and microscopic imaging [39, 40]. Scattering simply provides the colour and intensity contrast for different microscopic regions (e.g., crystallites) which are then imaged. Regarding the contrast mechanism, we consider that the expanding drops are ensembles of non-adsorbing, dielectric spheres. In general, their scattering properties depend on the magnitude of the scattering vector $q = (4\pi n/\lambda_0)\sin(\Theta/2)$, where $\lambda_0$ is the wavelength in vacuum, $n$ is the index of refraction of the solvent ($n_{H2O}$ =1.333), and $\Theta$ is the scattering angle. Images capture both coherent and incoherent, single and multiple scattering contributions (CS, IS, SS and MS, respectively) [41]. These vary with density and wavelength, and further depend on suspension structure. Additional experiments were conducted to carefully characterize the effects of a density-gradient and the melting transition on the optical properties (See Fig. S2a, S7 and S8 of the SI). From these, we deduce our drops to feature concentric layers of different scattering properties. Most prominent is a narrow, sheet of suitable density perpendicular to the radial gradient. Within and from it, we see very intense coherent MS caused by the 110-Bragg reflection and by incoherent scattering (Fig. S7a and b of the SI). Under monochromatic illumination, MS provides excellent imaging contrast, while the density dependence of the Bragg reflection allows precise localization of isopycnic-surfaces. At the same time, the increased reflectivity results in a drastic reduction of transmission [42] (Fig. S8b). The MS-sheets will hinder light from penetrating into the core underneath, which, despite increasing incoherent MS, is actually more transparent again. Outside the sheets, the suspension is fairly transparent and SS dominates, allowing for imaging of individual (200) and occasionally (211)-Bragg reflecting crystals (Fig. S6c and Video 1 in the SI). For the (110) reflection of the MS-sheet and the (200) reflections in the enveloping region, the local density is calculated as:

$$\rho = \frac{4n^3}{\lambda^3}\sin^3\left(\frac{\Theta_{110}}{2}\right) \text{ and } \rho = \frac{2n^3}{\lambda^3}\sin^3\left(\frac{\Theta_{200}}{2}\right) \quad (1).$$

At the melting transition radius, the mix of different scattering types changes again. This is seen in direct imaging as the absence of any Bragg reflections combined with a slightly increased turbidity, while in transmission, the boundary is identifiably as a change in the slope of the absorption gradient. Together, these effects allow a precise localization of the melting transition.

Regarding imaging, a combination of different illumination and observation arrangements was employed. A sketch of the optical set-up is given in Fig. 2. A more detailed drawing can be found in Fig. S4 of the SI. The setup allows for flexible switching between different imaging modes, which are presented by example images in Fig. 3a to c. We first consider the two alternative illumination paths.

Both in transmission and in the 90° scattering geometry, the sample (S) is illuminated by commercial white light sources ($L_1$, $L_2$) of 5000K [L, Avalight-DH-S; LS-0610025, Avantes B.V. Apeldoorn, NL]. These instruments deliver the light via a fibre-optic transmission line terminating in a collimation unit. Backside-illumination by $L_1$ is used for transmission imaging (TR). Here, we set the collimating unit to slightly divergent. The light impinges on a diffusor screen (Optolite™ HSR, Knight Optical, GB) to provide homogeneous diffuse background illumination at a width of 3 cm. To adjust image brightness at constant colour temperature, we inserted an attenuating neutral filter (NF; D1.0-D2.3, Edmund Optics, DE).

The illumination path coming from $L_2$ on the right side is used for imaging at $\Theta$ = 90° scattering angle. Here the collimation unit is set to deliver parallel white light at a beam width of $\approx$ 1.5 cm. A wheel with a neutral filter and several monochromatic band-pass filters ranging from deep blue to deep red (Edmund Optics, DE) allows flexible switching of the imaging mode. In white light Bragg-imaging (WB) we use the NF to adjust brightness. In monochromatic Bragg-imaging (MB) the selected $\lambda$ correspond to the following densities: $\lambda$ = 633 nm: $\rho$ = 31.54 µm$^{-3}$; $\lambda$ = 611 nm: $\rho$ = 35.07 µm$^{-3}$ orange; $\lambda$ = 590 nm: $\rho$ = 38.95 µm$^{-3}$; $\lambda$ = 547 nm: $\rho$ = 48.88 µm$^{-3}$; $\lambda$ = 514 nm: $\rho$ = 58.91 µm$^{-3}$; $\lambda$ = 488 nm: $\rho$ = 68.83 µm$^{-3}$). In several experiments reported here, we combined monochromatic illumination under 90° with transmission illumination (mixed mode, MM) to minimize overexposure effects. For drop shape checking, a sample illuminated by $L_2$ can alternatively be observed in transmission under $\Theta$ = 0° with the camera placed to the right side (not shown in Fig. 2).

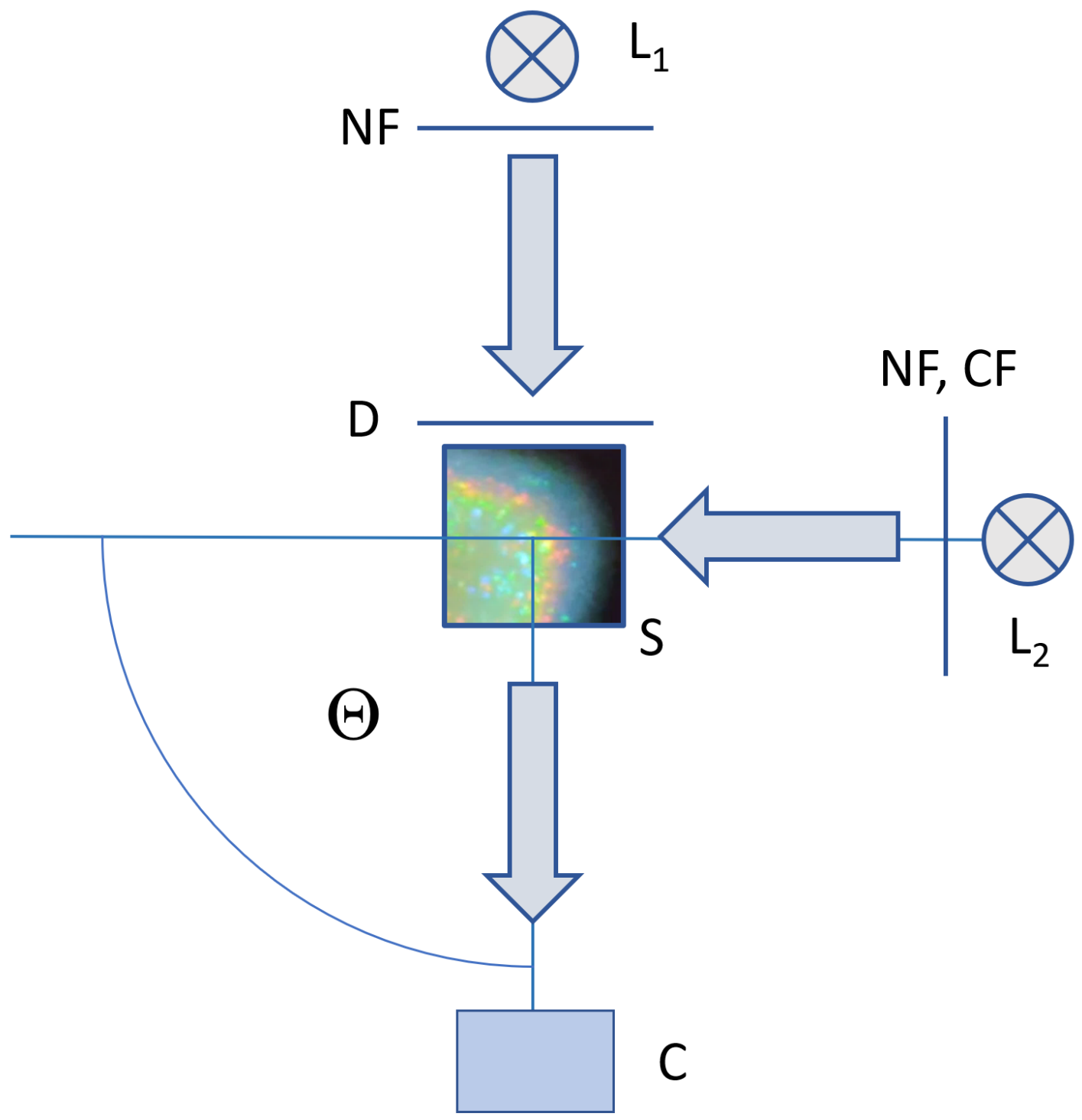

**Fig. 2 Optical set-up.** Top view with alternative illumination/detection paths (grey arrows). Optical elements are denoted by bold letters. $L_{1,2}$: white light sources; NF: neutral grey filter; D: diffusor; CF: narrowband filter; S: four-side-polished optical cell containing the expanding drop; C: camera. In the camera position sketched, we observe light from $L_1$ scattered under $\Theta$ = 90° (WB and MB mode) and/or transmitted light from $L_2$ (TR mode). The cell interior shows a typical WB image of a drop under white light illumination.

## 2.4 Imaging, image processing and analysis

All drops were observed in portrait mode with a 46Mpix consumer CMOS camera (C; Nikon D850) equipped with a belly and an inversely mounted f1.4, 50 mm lens. This resulted in an approximately 2:1 image on the sensor screen. ISO was set to 400 and exposure times were in the range of 1/50 s – 1/400 s. In MB image series with six colours, we have a minimum time resolution of $\Delta t_{min}$ = 10 s between subseries due to the time needed for filter switching. In each subseries, we switched colours from long to short wavelengths to counteract any timing-bias due to intermediate expansion. For a series of 6 colours, one white light and one transmission image, $\Delta t_{\mathrm{min}}$ = 30 s. The 14-bit raw images were saved to a computer in .nef-format. Examples of unprocessed images were shown in Fig. 1b. Subsequent image processing (centring, alignment, cropping, size calibration and colour temperature correction) was performed with ViewNX2 and/or NX Studio software (Nikon, JP) and the results stored as 16-bit .tiff-images. In each mode, the crystal spheres display a characteristic appearance and different quantities can be determined. Processed examples of images taken in TR, MM and WB mode are displayed in Fig. 3a, Fig. 3b and Fig. 3c, respectively. A set of characteristic observable extractable from the images is shown in Fig. 3d.

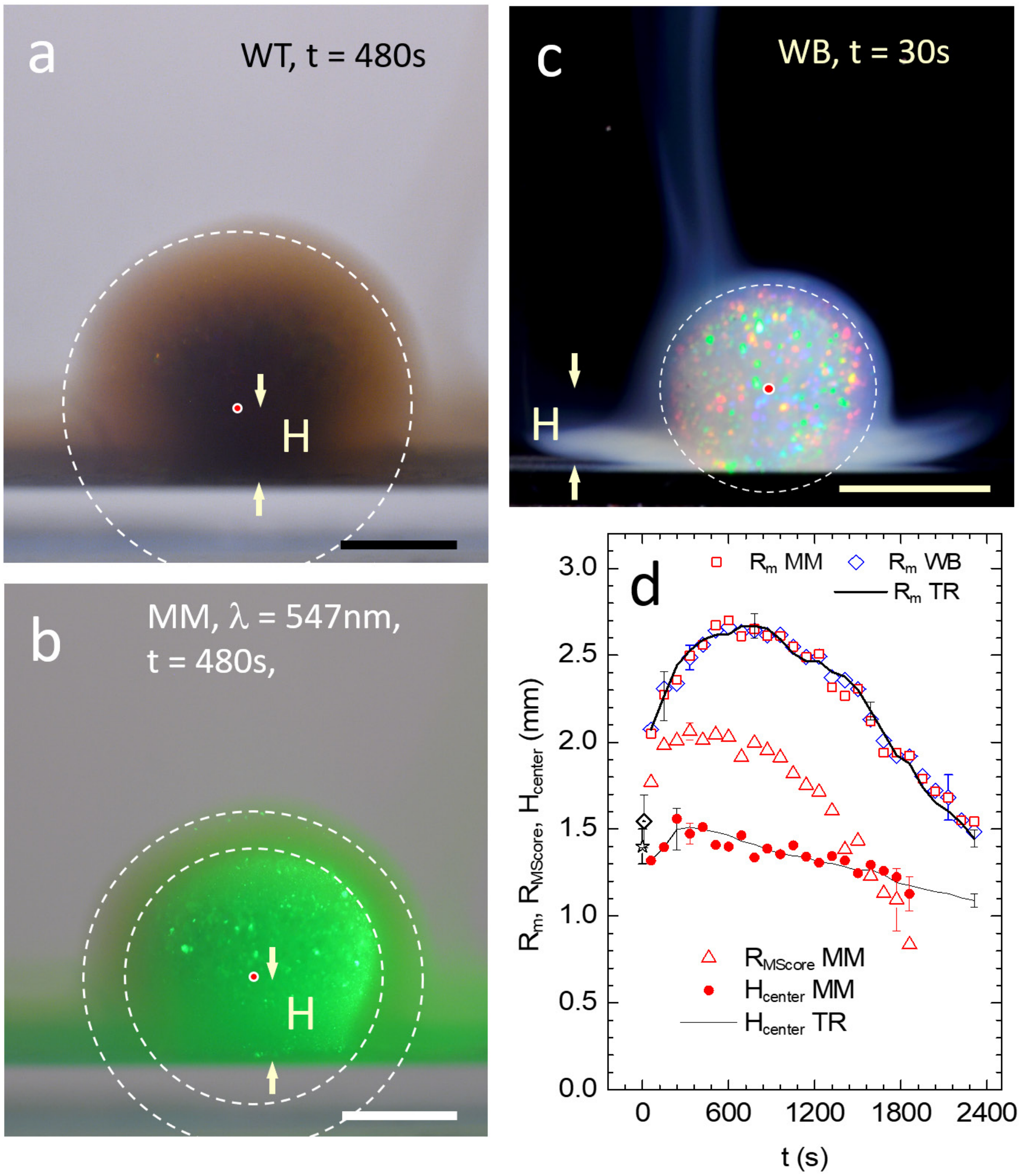

**Fig. 3 Examples of processed images from different illumination/observation modes and their evaluation.** The scale bars are 2 mm. a) TR: transmission-image taken at t = 480 s. The outer edge of the crystal is marked as a dashed ellipse. It corresponds to the melting density of $n_m$ = 15 µm$^{-3}$. A red dot marks the centre of the fitted contour and arrows mark its height, $H$, above ground. b) MM: Combined monochromatic 90°-Bragg- and white light transmission-image taken at t = 480 s. The red dot and the outer dashed ellipse denote - as before - the location of the crystal centre and its outer boundary, respectively. The inner dashed ellipse marks the MS-surface for λ = 547 nm., i.e., the location of a concentric shell of density $n$ = 48 µm$^{-3}$. With an initial density of $\rho_0$ = 110 µm$^{-3}$ the probed density ratio is $\rho_m / \rho_0$ = 0.13. c) WB: white light Bragg-image taken shortly after touch down at $t$ = 30 s. The MS-core region, visible as a red-scattering band, nearly reaches out to the crystal/fluid border (dashed line). The centre of mass is still very close to its initial height. Note the blueish hue of the fluid phase, embedding the crystalline phase and partially sedimenting to the cell bottom. d) Comparison of representative results from images of one spherical drop, captured in different modes, as is indicated in the legend. $R_m$: equivalent radius of the crystalline region; $R_{MScore}$: wave length dependent location of the 1st order Bragg multiple-scattering shell; $H_{center}$: height above ground of the fitted ellipses. Also shown are the equivalent radius in air (star) and the drop radius at $t$ = 5 s after touchdown, as inferred from combining WB and TR results (dotted diamond).

In the TR image, one notes a gradual increase of extinction towards the centre. This projection effect results from the spherical contour of the MS-layers. Additional smoothing is provided by their wave length dependent location and the overall strong extinction by incoherent MS. Figure 3b shows a typical MM-image. Here, the drop-core appears as a compact, uniformly coloured sphere. Its fine surface-texture originates from individual reflections of small crystallites embedded in a uniform MS-background. This (110)-MS sphere shows an excellent contrast to the surrounding, crystalline region. It can be precisely localized and followed in time (cf. Video 2). Remarkably, under variation of the illumination wavelength, we observe different locations for differently coloured MS-shells relating to different isopycnic surfaces. This allows using eqn (1) to infer the corresponding density profile. These MS-shells are embedded in a more transparent outer region with only few reflections, in which individual crystallites can be distinguished.

In the early-stage WB image of Fig. 3c, the overall impression is a turbid, milky-white core dotted with individual, coloured reflections. We attribute this to strong incoherent MS at all illuminating wavelengths. Note that initially, differently coloured reflections are observable over the whole drop (cf. Video 1), which can be attributed to additional Bragg scattering of individual crystallites. After about a minute, however, a concentric colour banding appears, presenting a characteristic radial sequence of dominant colour, which relates to the MS-core (see also in Fig. 1b). As the MS-core retreats at later stages, it gets surrounded by an extended transparent crystalline region. This region shows much weaker IS and (200) Bragg reflections become nicely visible (Fig. 5b, below). In principle, eqn (1) should be applicable to these reflections, given that the scattered wavelength can be identified. We tried to extract the latter from the corresponding RGB readings in WB images using a recently proposed protocol [43]. Yet, this approach failed for the monochromatic (200) Bragg reflections of individual crystals. The scattered wavelengths are lying outside the RGB-triangle in the IECE colour chart [44, 45]. This leads to hue-dependent calibration issues and the RGB readings cannot not be unequivocally assigned to individual scattered wavelengths (See Fig. S9 in the SI). This prevented density profiling in the transparent outer regions.

General shape fitting, determination of the centre of mass, object tracking for drifting crystals and brightness analysis for the TR and MM images used standard Image-j routines [ImageJ; http://imagej.nih.gov/ij] as well as home-written algorithms. Specifically, we fitted ellipses of horizontal and vertical semi-axes, $a$ and $b,$ respectively, to the projections of the crystal-fluid boundary for all three modes. The boundary is shown in Fig. 3a-c as a dashed line. In TR, we located it from the change in radial slope of the transmission signal, in WB and MM, we used the scattering contrast provided by the outer border of Bragg reflections and the simultaneous increase of CS in the fluid phase. In MM, we further fitted an ellipse to the outline of the central MS core in dependence on wavelength using brightness contrast. Its contour is shown in Fig. 3b as the inner dashed ellipse. From the semi-axes of fitted ellipses, we calculated the radii of equivalent spheres as: $R_{\mathrm{i}} = (a^2b)^{(1/3)}$. Note that for sufficiently small eccentricities, $a \approx b$, and the centre-to-rim distances approaches the radius $R_{\mathrm{i}}$ of the equivalent sphere. Irrespective of imaging mode, the outermost ellipse yields the equivalent radius of the crystalline region, i.e., the melting radius, $R_{\mathrm{m}}$. In MM mode, we additionally determine the wave length dependent equivalent radius of the $1_{\mathrm{st}}$ order Bragg multiple-scattering shell, $R_{\mathrm{MScore}}$. Further, the centres of the ellipses yielded the heights, $H_{\mathrm{center}}$, of the drop centre-of-mass above ground. Figure 3d shows the evolution of these observables for the example drop shown in Figs. 3a to c. The error bars are estimates based on deviation from rotation symmetry, contrast issues and interfacial smoothness. Since the crystallite number decreases while the crystal sphere size increases, the assignment of an outer boundary from WB gets somewhat less certain toward late stages. The same applies to the monochrome MM images, albeit at generally smaller uncertainties. Irrespective of imaging mode, however, data on $R_{\mathrm{m}}$, $R_{\mathrm{MScore}}$ and $H_{\mathrm{center}}$ agree well within experimental uncertainty. This agreement worsens only slightly for the two shortest wavelengths used in MM. There, incoherent MS emerging from the corresponding blue and cyan (110) Bragg scattering shell is pronounced enough to slightly blur its contour.

## 3 Results

### 3.1 General scenario

In air, the drop has an equivalent radius of about $R_{\mathrm{air}}$ = 1400 µm. Immediately after entering the water, it starts expanding and re-crystallizing. Mechanical stabilization starts upon contact between individual crystallites. On average, the shape became stable at $t \approx 5$ s with a corresponding average radius of $<R_{\mathrm{m}}(t \approx 5\ \mathrm{s})>$ = (1550$\pm$150) µm. After some ten minutes, the crystalline part of the drop reaches its maximum radial extension of about $R_{\mathrm{m}} \approx 2500$ µm. Thus, drops expand while they are diluting simultaneously. The broadening of the inward shifting coloured banding in the WB images indicates a decreasing slope of the density profile as well as an ever decreasing values for the central density. Thus, we observe two continuous but counterpropagating modes of motion for the observables. Initially, expansion dominates the location of a certain density or lattice constant, later on dilution takes over but no stationary state is reached.

A measurement focusing on the short-time crystal sphere expansion is shown in Fig. 4, where we display the expansion-curves for $R_{\mathrm{m}}$ and for the outer rim of the reddish band $R_{\mathrm{red\ stripe}}$ seen in Fig. 1b and 3c. (The shown equivalent radii were derived from fitted ellipses as described above) In this double logarithmic plot, the initial increase is seen to follow a power law, with a significant exception for the very first data point. While at this time ($t$ = 5 s) the crystal shape has already stabilized to a sphere, the optical appearance is misleading, in that it suggests completed solidification. In fact, despite their optically compact appearance, polycrystalline materials may remain semi-solid for some further time until full rigidity is obtained after $t \approx 5$-8 s [46]. Therefore in Fig. 4, we can safely assume completed crystallization only for the power-law increase seen between 10 s and 200 s. After reaching a maximum, the crystalline region shrinks due to inward melting, as does the core region due to ongoing dilution. The latter disappears after some 25-30 min, the last outer crystals melt after some 30-45 min slightly depending on drop history and shape.

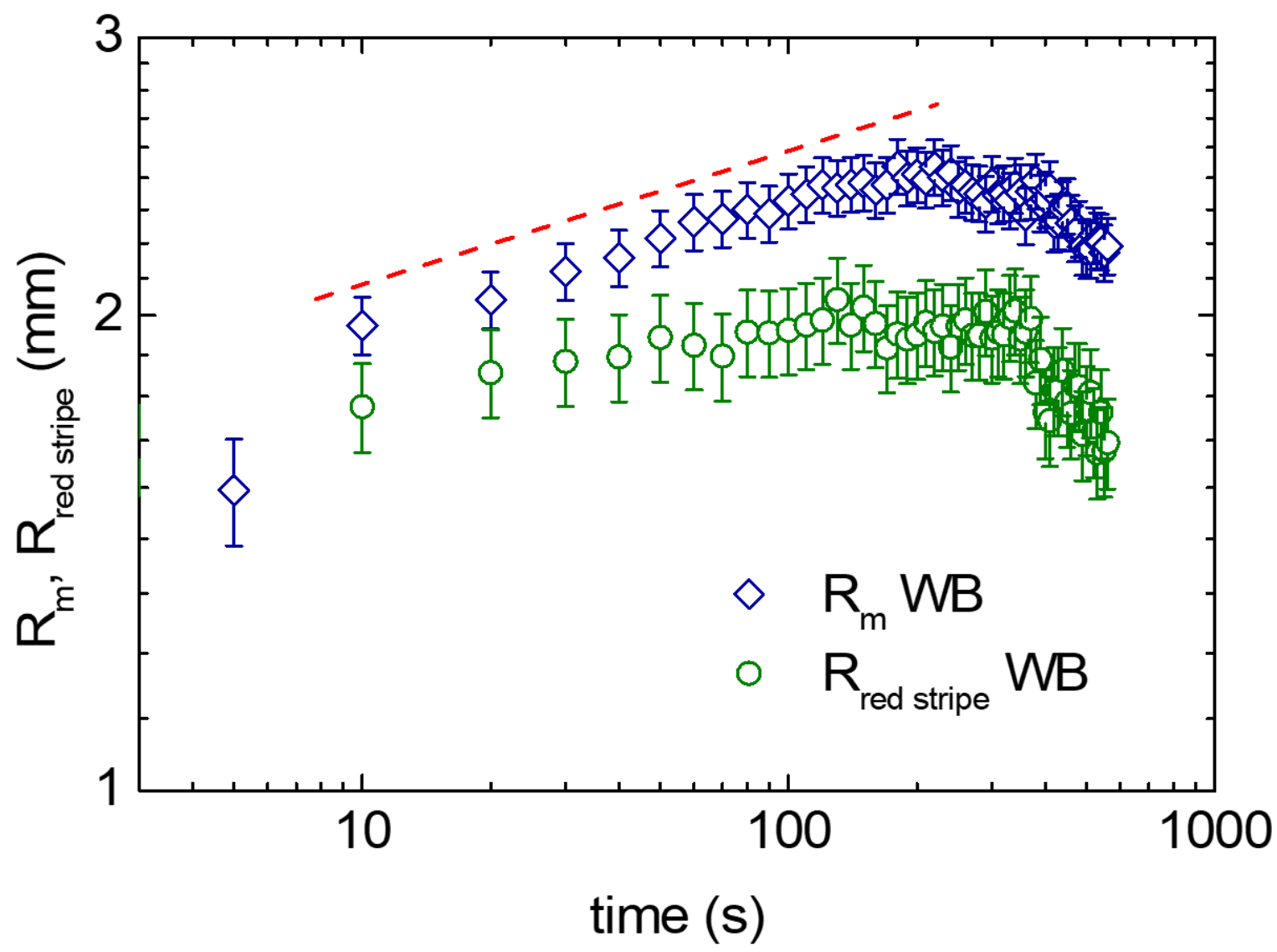

**Fig. 4 Short time expansion.** Double-log plot of deduced radii for a single run versus time. Symbols as indicated in the key. The red dashed line shows a power-law behaviour for comparison. Note the deviation of the first $R_m$ data point.

3.2 Expansion and dilution on the level of individual crystallites

Immediately after solidification, we observe a fine-grained polycrystalline solid encapsulated by a thin fluid layer (Fig. 3c). Throughout drop evolution, individual crystals show an outward drift and an increase in size, while their persistent and directed colour-changes demonstrate their ongoing dilution. This is illustrated in Fig. 5 for two runs starting at $t_0$ = 150 s and $t_0$ = 2820 s.

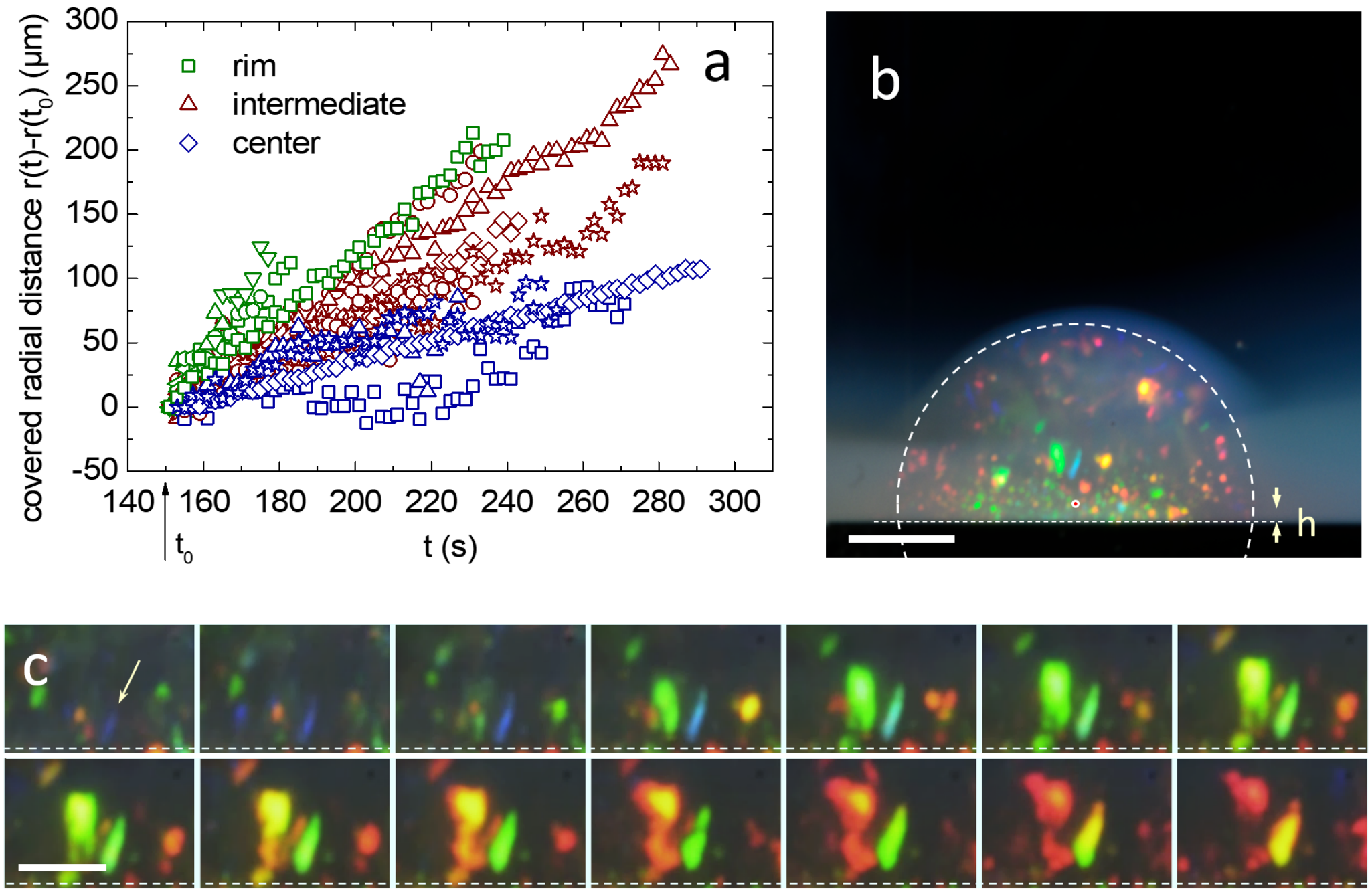

**Fig. 5 Crystal drift and expansion.** a) Covered distance in radial direction $r(t)$-$r(t_0)$ versus time $t$ for several individual crystallites discernible at $t_0$ = 150 s. The colour coding indicates the crystallite position relative to the drop centre at $t_0$. b) Late-stage drop imaged in WB at $t$ = 3000 s. The scale bar is 1 mm. c) cropped WB images taken from $t_0$ = 2820 s to $t$ = 3660 s in intervals of 60 s. The scale bar is 500 µm. The dashed line is a guide to the eye, marking a constant height above ground of 250µm. The arrow marks a faint blue crystal at the start of this series, which evolves in both size and colour. The long duration of the green scattering stage is an illusion of the camera's and the eye's RGB sensitivity. In fact, the scattered wavelength changes continuously.

Individual crystallites could be followed over 30 s to 120 s, during which they typically showed constant radial velocities ranging between 0.2 µms$^{-1}$ in the central region to 15 µms$^{-1}$ close to the drop rim. This spatial variation is shown in Fig. 5a. The observed correlation of drift velocities to crystal location is attributed to projection effects, as centrally seen crystallites move preferentially towards the camera, while rim crystallites drift preferentially perpendicular to that direction. In Fig. 5b, we show an evolved drop with a shrunken core. In Fig. 5c, we display a series of cropped images focusing on the evolution of an individual crystal close to the core-rim. As it drifts slowly outward, the initially small dark blue crystal expands considerable and continuously changes its Bragg-reflected colour. The crystallite to its immediate right even shows colour banding within its interior. Both directly visualize the ongoing interplay of drop expansion and dilution.

## 3.3 Size evolution of the crystalline part of the drop.

A typical MM time series for $\lambda$ = 611 nm is shown in Fig. 6a. (For large images and the $\lambda$ = 590 nm series, see Fig. S9 in the SI). Results for $R_m$ from MM at different illumination wavelengths are compared in Fig. 6b to results from TR. The overall agreement is very good, except for the shortest wavelengths, which at later stages show a slight systematic deviation towards larger values. This is attributed to blurring caused by incoherent MS. In Fig. 6c, we omit these two data sets for small $\lambda$ and display the volume of the crystalline part of the drop, calculated as: $V_m = (4\pi/3)\, R_m^3$. Data from both observation modes coincide nicely within experimental scatter. The initially fast volume increase gradually slows and past the maximum expansion at $t \approx$ 700 s, the crystal sphere volume decreases. The straight, dashed line is a guide to the eye. Over more than 1000 s, the decrease is nearly linear. The observed late stage slowing may relate to the crystalline region being meanwhile immersed in a rather concentrated fluid environment (see last images in Fig. 6a).

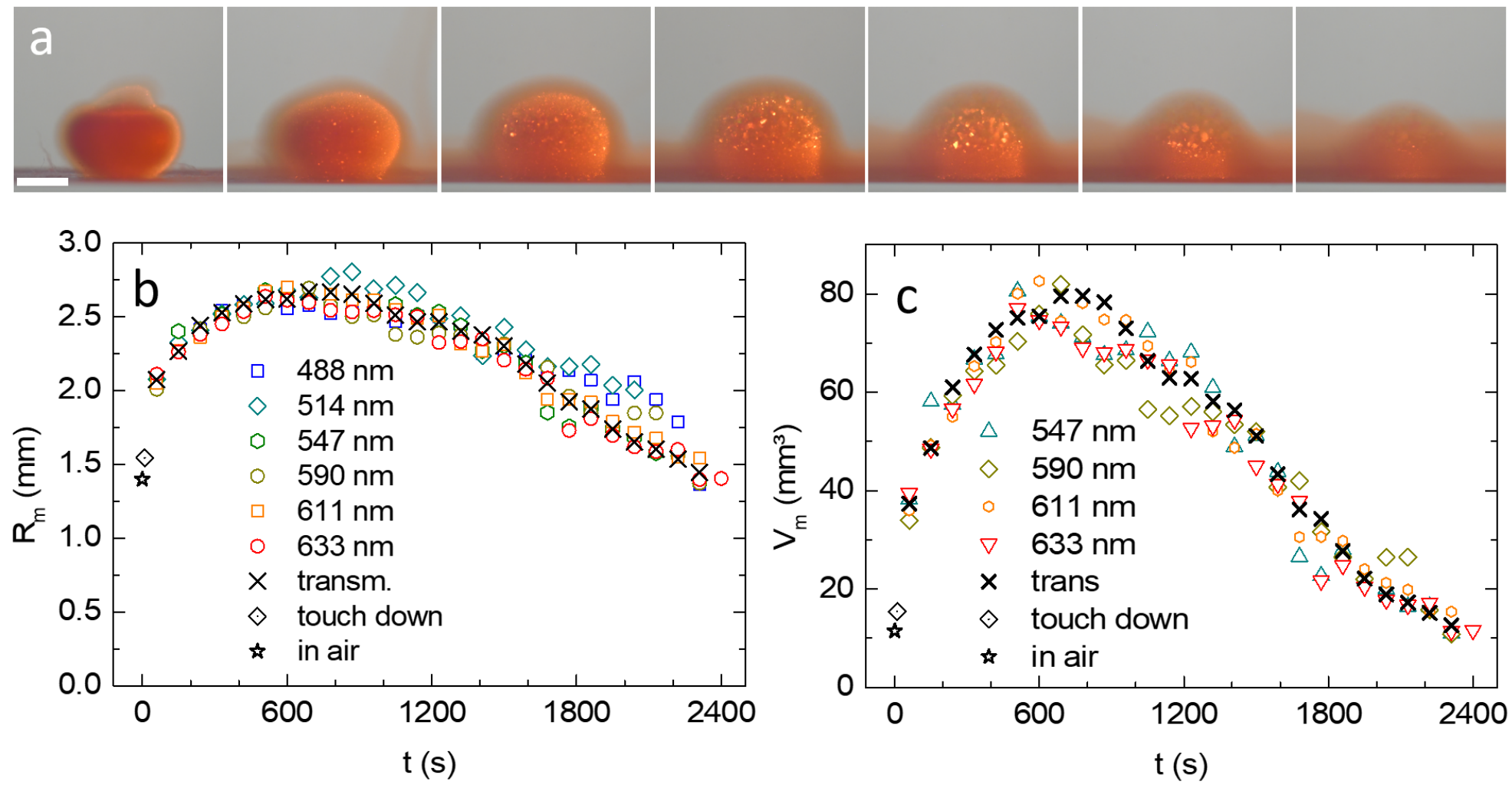

**Fig. 6 Expansion curves for the crystalline part of the drop.** a) MM time series taken at $\lambda$ = 611 nm. The scale bar is 2 mm. b) Equivalent crystal sphere radii from MM series at different wavelengths as indicated in dependence on time after release as obtained. Also shown are the equivalent radius in air (star) and the equivalent drop radius at $t$ = 5 s after touchdown (dotted diamond) as inferred from combining WB and TR results. c) Equivalent crystal volumes in dependence on time after release for wavelengths selected for low systematic uncertainty in radius determination. Symbols as before.

## 3.4 Density profiling

A colour series of MM images taken at $t$ = 480 s is displayed in Fig. 7a. Most strikingly, we observe a concentric nesting of the thin, isopycnic MS-shells, vividly Bragg scattering a certain wavelength but obviously transparent to all others. The derived $R_{MScore}(t)$ are shown in Fig. 7b. As the core regions expand and shrink again, the wavelength dependent shell-radii keep their sequence, i.e. the largest densities are consistently found in the innermost shell. This nesting qualitatively confirms the intuitively expected radial gradient in density and allows for density profiling. Figure 7c shows the derived radial density profiles for different times. At all times, we see a radially decreasing density profile. Initially, the profile follows the general expansion and shifts rightward. This trend is later reversed due to core shrinkage. Most importantly, the initial density profiles are rather steep, but flatten continuously with time.

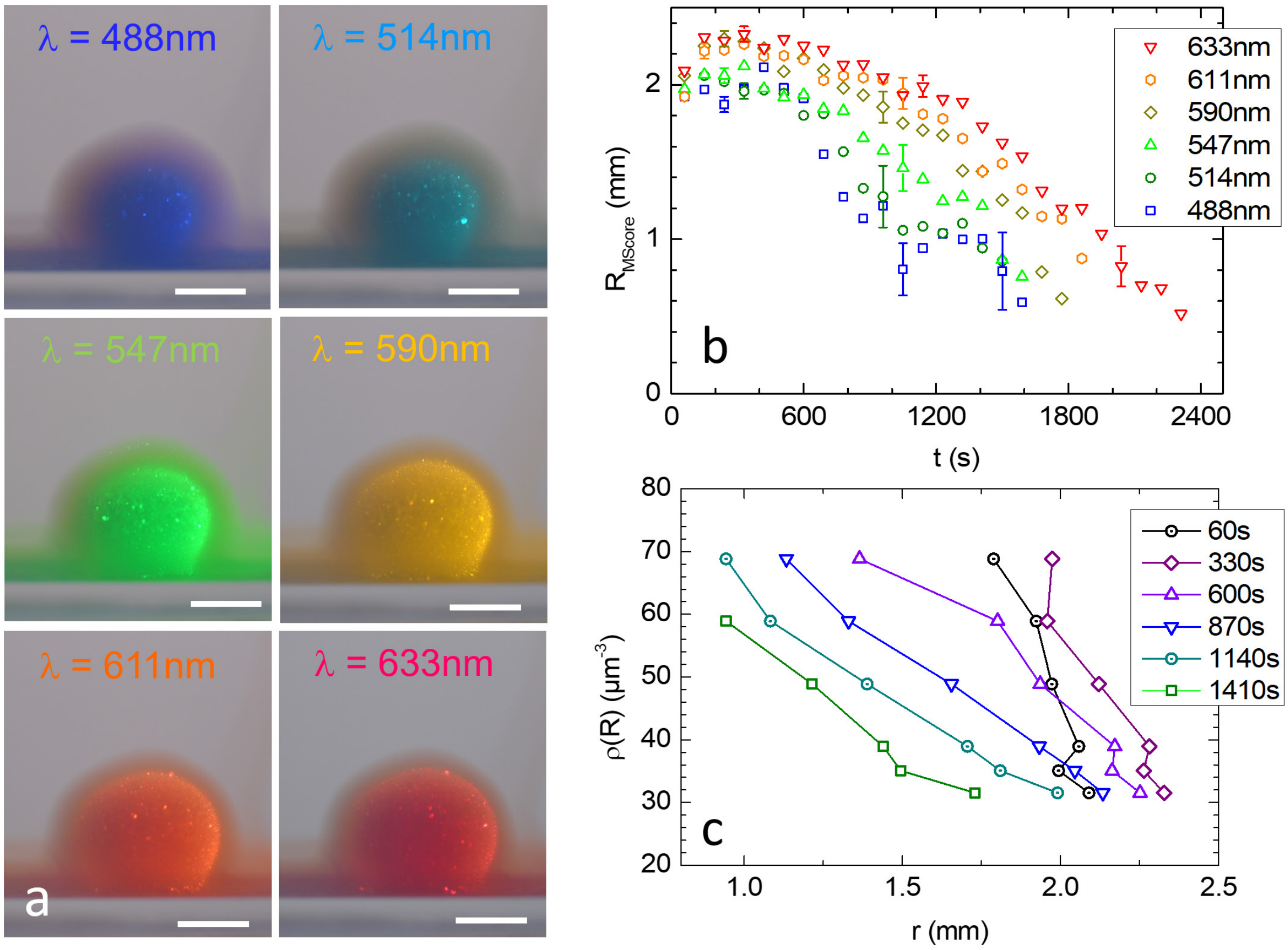

**Fig. 7 Evolution of the crystal sphere core – density profiles**. a) Processed MM images recorded at $t$ = 480 s for different wavelengths as indicated in the key, ranging from dark blue to dark red and corresponding to densities from $\rho_m$ = 31.54 µm$^{-3}$ to $\rho$ = 68.83 µm$^{-3}$, respectively. Note the nested structure of shells with different densities, i.e. $R_{488} < R_{633} < R_m$. Scale bars 2 mm. b) Equivalent MS-core radii obtained in MM. Note the increase in uncertainty for the shortest wavelengths due to blurring by incoherent MS. c) Density profiles as obtained from the radial positions of MS-core scattering using eqn (1) for different times indicated in the key. Note the initial right shift and the gradual decrease in profile steepness.

## 4 Modelling

### 4.1 Dynamical density functional theory for the isothermal expansion of a spherical crystallite

We now turn to a theoretical description of the crystal expansion process. We use the framework of classical dynamical density functional theory (DDFT) which provides an ideal microscopic framework for expanding solids [30, 47]. We apply it here in a simple way to describe the isothermal expansion of a three-dimensional spherical colloidal crystallite in the bulk in the absence of any boundary and solvent flow. The crystal sphere dilutes by expansion and simultaneously melts at its contour. The remaining crystalline portion decreases. We consider a coarse-grained inhomogeneous time-dependent density field with radial symmetry, $\rho(r, t)$, where $r > 0$ denotes the distance from the centre of the crystallite.

The initial density distribution is modelled as a steep $\tanh$-profile:

$$\rho(r,t=0)=\rho_0\left(1-\tanh\left(\left(r-r_0\right)/\xi\right)\right) \qquad (2),$$

where $\rho_0$ is the initial crystalline density and $r_0$ is the initial radius of the spherical crystallite, while $\xi$ describes the initial width of the solid-fluid interface. For the dynamical evolution of the density field $\rho(r, t)$, we follow [30] and write it as

$$\frac{\partial\rho(\vec{r},t)}{\partial t}=\nabla\cdot\left(\frac{D}{k_B T}\rho(\vec{r},t)\nabla\frac{\delta F\left[\rho\right]}{\delta\rho(\vec{r},t)}\right) \qquad (3),$$

where $D$ is the diffusion coefficient of the colloidal particles and $k_B T$ is the thermal energy with system temperature $T$. The free energy density functional $F[\rho]$ is given within the local-density approximation (LDA) as

$$F\left[\rho\right]=\int d^3r\left(k_B T\rho(\vec{r},t)\left(\ln\left(\Lambda^3\rho(\vec{r},t)\right)-1\right)+f_{exc}\left(\rho(\vec{r},t),T\right)\right) \qquad (4).$$

Here, $\Lambda$ is the thermal wavelength (for dimensional reasons only) and $f_{exc}(\rho(r, t),T)$ the excess free energy density per volume of a bulk system at temperature $T$ and homogeneous density $\rho$ in the solid or fluid phase. We approximate $f_{exc}(\rho(r, T)$ by the coarse-grained potential energy of a bcc-crystal. We use a lattice sum at zero macro-ion temperature with the DLVO-like Yukawa pair potential [48]

$$V(R)=\frac{Z_\sigma^{\ 2}}{4\pi\varepsilon_0\varepsilon_r}\frac{\exp(-\kappa R)}{R} \qquad (5),$$

where $R$ is the distance between two colloids. In the specific case of a bcc-lattice with a lattice constant $a_0$ at $t$ = 0, $R$ = $\left(\sqrt{3}/2\right)a_0$ for the nearest neighbours and accordingly for the next nearest neighbours. $Z_\sigma$ is the effective macroion charge, $\varepsilon_0$ the dielectric permittivity of vacuum, $\varepsilon_r$ = 80 the dielectric constant of the solvent, and $\kappa$ the inverse Debye-Hückel screening length [49]. We express the excess free energy approximately as

$$f_{exc}(\rho,T)\approx\frac{1}{a^3}\sum_{\vec{R}_N}V\left(\left|\vec{R}_N\right|\right) \qquad (6).$$

The density dependence enters via the bcc-lattice constant $a = (2/\rho)^{1/3}$. The sum in eqn (6) extends over all lattice vectors $\vec{R}_N$ of the three-dimensional bcc-lattice. In radial symmetry, eqn (3) reads:

$$\frac{\partial\rho(r,t)}{\partial t}=\frac{1}{r^2}\frac{\partial}{\partial r}\left\{r^2 D\left(\frac{\partial\rho(r,t)}{\partial r}+\rho(r,t)\frac{\partial}{\partial r}\frac{\partial f_{exc}(\rho,T)}{\partial\rho}\right)\right\} \qquad (7),$$

which we solve numerically with the initial condition at t = 0 given by eqn (2) with $\xi \rightarrow 0$, i.e., a rectangular function corresponding to no initial solid-fluid interface. Thus, the time-dependent crystalline volume during the expansion is given by

$$V_C(t)=\frac{4\pi}{3}r_C^3(t) \qquad (8),$$

where $r_C(t)$ is the radial distance of an isopycnic spherical surface of density $\rho_C$, for instance the melting density $\rho_m$, which is known for a three-dimensional Yukawa system [50, 51, 52, 53]. I.e., $r_m(t)$ is determined by the condition:

$$\rho\left(r_m(t), t\right) \equiv \rho_m \qquad (9).$$

In the calculations, $\rho_C$ can further be viewed as a flexible parameter that determines the crystalline volume enclosed by an isopycnic surface corresponding to any given bcc-lattice constant. Results for any chosen $\rho_C$ can thus be directly compared to our experimental findings. Our numerical solutions of eqn (7) are shown in Fig. 8.

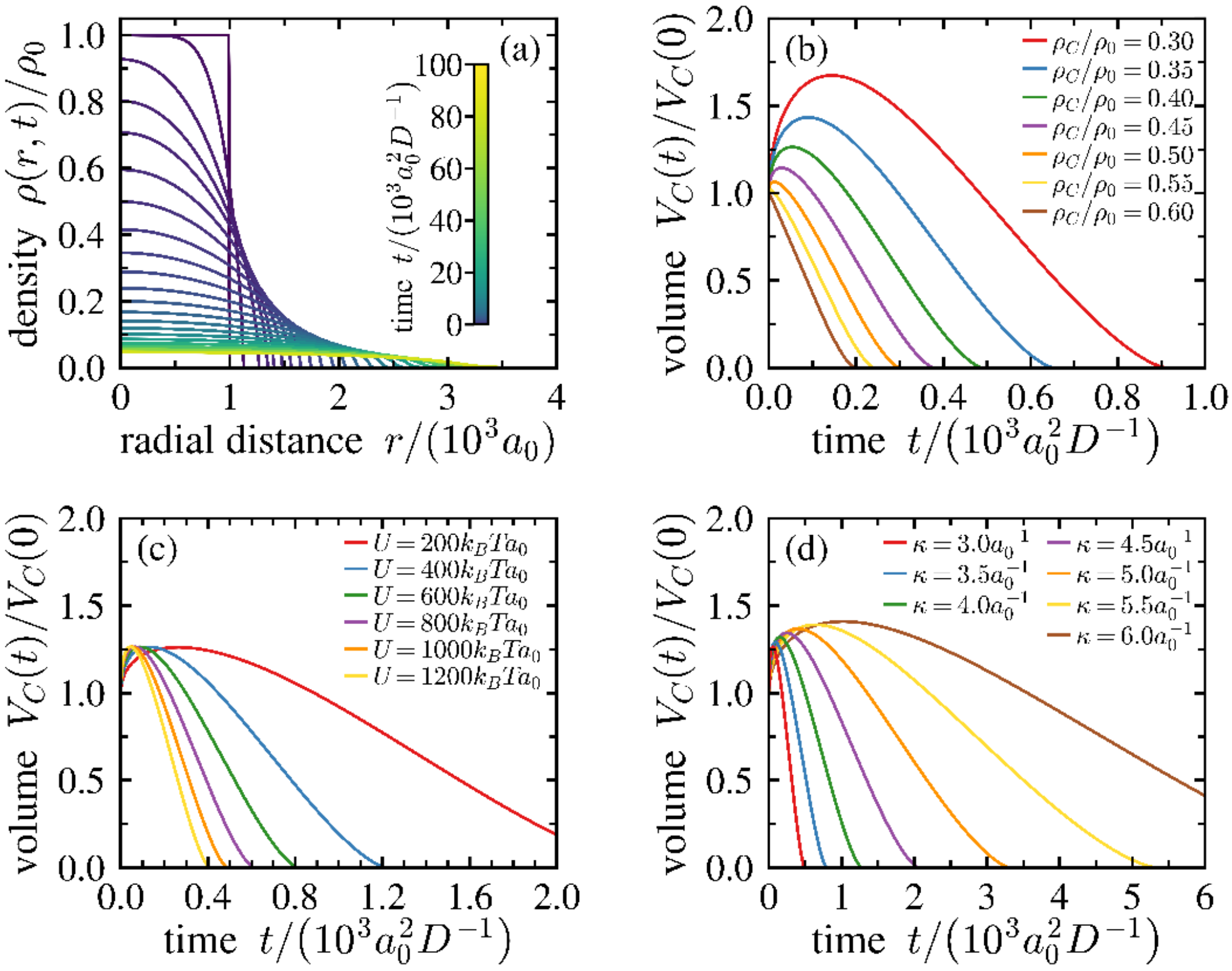

**Fig. 8 Numerical solution of the dynamical density functional theory equation.** (a) Density evolution of the crystallite. During the crystal expansion, the local density decreases, which leads to melting. The initial size of the crystallite is $r_0 = 10^3 a_0$, the inverse Debye-Hückel screening length is $\kappa = 3a_0^{-1}$ and the strength of the Yukawa pair potential is $U = 10^3 k_B T a_0$, where $a_0$ is the initial bcc-lattice constant given by $a_0 = (2/\rho_0)^{1/3}$. The sum of eqn (6) is approximated by truncating after the fourth nearest neighbours. (b) Time dependence of the crystalline volume $V_C(t)$ (i.e., the crystalline volume enclosed by an isopycnic shell of density $\rho_C$) for various densities $\rho_C$ as indicated in the key. Here, $\kappa = 3a_0^{-1}$ and $U = 10^3 k_B T a_0$. For $\rho_C \leq 0.55\ \rho_0$, the crystallite initially expands before its volume decreases linearly in time over an extended period. (c) The same for $\rho_C/\rho_0 = 0.40$, $\kappa = 3a_0^{-1}$, and different Yukawa repulsion strengths, $U$, as indicated in the key. $U$ has little influence on the functional form of the curve but leads to a stretching of the time scale of expansion. (d) The same for $\rho_C/\rho_0 = 0.40$, $U = 10^3 k_B T a_0$, and different $\kappa$ as indicated in the key. An increased $\kappa$ leads to a slight increase of the curve maximum and a stretching of time scales.

Suitable start-parameters were chosen close to those in the experiment and are provided in the legend and in the captions. We find a continuously flattening density profile in Fig. 8a. Simultaneously, the central density decreases (cf. Fig. S11 of the SI). Note that at large times, the crystal core has developed a near constant density, and a significant slope is observable only at an increasing radial distance. For the core region, this implies a homogeneous and isotropic expansion at late times. While the decrease in central density with time, as well as the radial decrease of density can be intuited, this specific prediction for the screened Yukawa interaction cannot and further differs markedly from the results obtained for plasma crystals [27].

Each numerical run at given start parameters yields a set of density profiles, from which the characteristic expansion curves for any chosen $\rho_C/\rho_0$ can be derived. Curves for different $\rho_C$ generated from the run shown in Fig. 8a are shown in Fig. 8b. For small $\rho_C$, we have only expansion. For density ratios $\rho_C \leq 0.55\ \rho_0$, an initial expansion precedes the dynamical decrease of the crystalline volume $V_C(t)$. There, the volume evolution features a pronounced maximum followed by a stretched, almost linear decrease as a function of time $t$. Note the pronounced increase of the maximum volume for decreasing density ratios $\rho_C/\rho_0$. Performing repeated calculations under systematic variation of parameters then shows the dependence of expansion curve shape and extension on these. Figure 1c and d demonstrate that the strength of the Yukawa potential $U = Z_\sigma^2 e^2/4\pi\varepsilon_0\varepsilon$ has practically none, and the inverse screening length has only moderate influence on the maximal expansion, but both influence the temporal evolution of the volume for a given density $\rho_m$.

Finally, we remark that the theory is simple in the sense, that it involves only the density as a dynamical order parameter and basically uses a local density approximation for the density functional. Non-local extensions with a full density field which is strongly peaked on the colloidal length scale would imply much larger numerical effort. The zero-temperature lattice sum for the Yukawa system is another approximation that may lead to deviations from the behaviour of an actual Brownian Yukawa system. Moreover, we assumed that the mobility of the particles is the same in both the crystal and the fluid phase. This is clearly an approximation that induces a smearing of the density profiles [54]. However, our simpler version already yields a specific prediction for the expansion of colloidal crystals as well as valuable information on the influence of important experimental parameters on the expansion, which cannot be controlled well in plasma experiments. By contrast, they are readily adjusted in the present study via $CO_2$ content and salt concentration [35].

## 5 Discussion

Thus far, we reported the fabrication of initially highly concentrated drops of suspensions made of strongly repulsive spheres in a sphere-free environment. This allowed for the first time to study the isotropic expansion of a polycrystalline solid in three dimensions. In the transparent parts of the drops, we used white light Bragg imaging to monitor the drift motion and expansion of individual, Bragg reflecting crystallites. Bragg imaging and complementary transmission imaging gave consistent results for the crystal sphere radii with high spatial resolution. We could overcome the limitations of white light Bragg imaging using monochromatic illumination Bragg imaging. Exploiting the peculiarities of coherent multiple scattering at Bragg reflections then allowed probing the locations of certain densities within the turbid drop core as well as density profiling.

Figure 9 compares the measured expansion curves to our theoretical expectations. To obtain the best overall least squares fits, we here varied the ratios $\rho_C / \rho_0$ as indicated in the key. The other input parameters were fixed to values corresponding to the experimental ones: $Z$ = 365; κ = 3.5; $D$ = $4D_0$ (except for $V_m$, where $D = 7.6D_0$); $\rho_0$ = 110 µm$^{-3}$; and $R_0$ = 1.40 mm. The fits capture the general curve shape very well. On the one hand, this demonstrates that our novel experimental approach can provide successful and reproducible access to density profiling in turbid crystalline media. Moreover, the remaining statistical and systematic uncertainties (c.f. Figs 3d, 6c and 7b) are small enough allowing for a quantitative comparison to theoretical predictions. On the other hand, Fig. 9 clearly shows that, in spite of the simplicity of the theoretically considered situation, our modelling approach is sufficiently sophisticated to yield an excellent and nearly quantitative description.

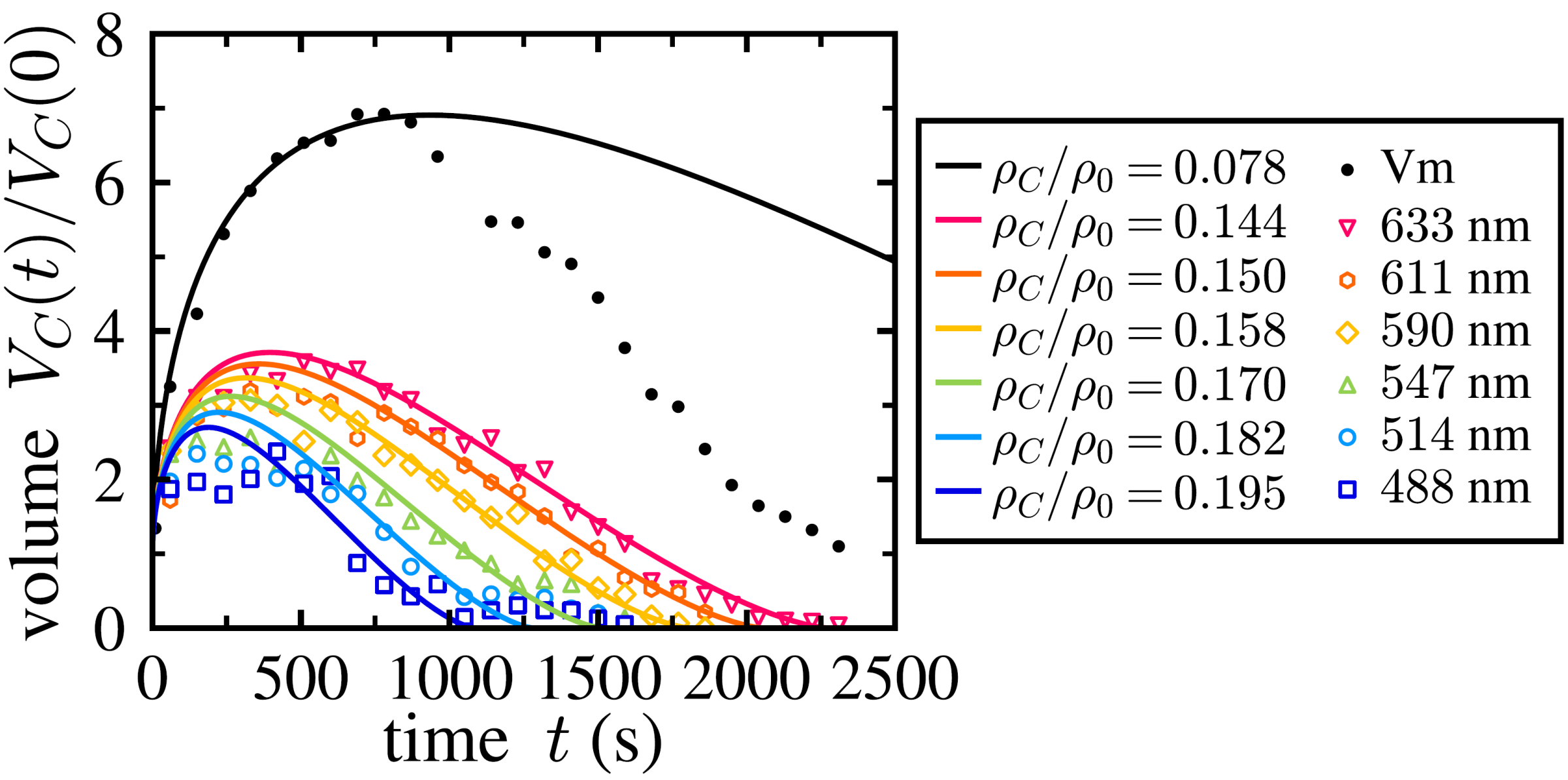

**Fig. 9 Temporal evolution of volumes enclosed by isopycnic surfaces.** Data are normalized to the initial drop volume. Symbols represent measured volumes; solid lines are least squares fits of the theoretical model using the ratio $\rho_C / \rho_0$ as only free parameter. Probing wavelengths and density ratios used in the fits are colour coded as indicated in the key. Note the overall very good description of experimental curves. Note further i) the pronounced deviation past the maximum of the melting density, ii) the flatter-than-expected density evolution at early times for the innermost densities, and iii) The discrepancy between the experimentally probed densities and the density ratios used in the fit.

As compared to previous studies on isotropic, three-dimensional expansion, polycrystalline aqueous charged sphere suspensions show a behaviour midway between that of Coulomb balls and Yukawa balls [27, 28, 29]. The former revealed a strictly flat density profile, while the latter produced strongly peaked profiles. Concerning two-dimensional experiments, our findings bear resemblance to those of Tanaka et al., who studied short range repulsive spheres in slit confinement. They found a continuously decreasing core density, albeit with only a weak density gradient within the crystal. They also reported the crystalline core embedded in a widening interfacial boundary region towards the fluid phase. In the present experiments, the innermost density profile was not (yet) accessible, but the density profiles observed roughly halfway remained for extended times while flattening continuously. Further, in the outer, transparent region, we could show clear evidence of a persistent radial density gradient. Crystal spheres made of charged colloidal spheres therefore provide a valuable platform for future systematic experiments. Particles of different sizes and charge densities are at hand, which will facilitate tuning of the time-scales and the repulsive strength. Using a polymer solution as solvent for the suspension and/or the receiving bath may further allow investigations of the effects of (depletion-) attraction and/or facilitate damping of the expansion by introducing an outer osmotic pressure. Experiments with drops of lower start density are under way, which should yield better access to the innermost drop regions. Experiments using X-ray tomography appear to be a possible alternative approach to monitor the density evolution throughout the complete drop.

Figure 9 also shows a few systematic deviations between experiment and theoretical model. There appears to be a mismatch between the density ratios, $\rho_C/\rho_0$ probed in the experiments starting at t = 60 s and the densities used in obtaining the overall best fits. Interestingly, also the other two discrepancies are restricted to certain time intervals. In these, the corresponding statistical and systematic errors considered in Fig. 3d and 6c are rather small. The observed deviations are thus both characteristic and significant. Further, in both cases, these discrepancies occur at characteristic times. In particular, at early times and for the largest densities probed, the measured and expected core-volume expansion curves appear flatter than expected, showing only weak maxima, while at late times experiment and theory are in very good agreement. Conversely, the measured crystal sphere volume, $V_{\mathrm{m}}$, follows the expectations closely at early times, while from some 800 s onward, it deviates towards smaller than expected volumes. None of these disparities can be attributed to the level of sophistication used in our model. In fact, modelling the expanding drops as isotropically expanding crystals made of electrostatically repulsive Brownian spheres and considering only due to density reduction at otherwise constant interaction parameters performs well enough to yield an overall excellent description of the experimental observations. Rather, the following discussion traces the mismatch back to violations of the required constant boundary conditions, which occur on the experimental side.

First, at the start of the expansion experiments the basic condition of an overall crystalline drop is only given for the model. On the experimental side, the drops are shear-molten by their extrusion and the impact on/in the water. Start of re-crystallization occurs at $t \approx 0.3$ s, but complete solidification takes considerably longer. Therefore, the drops initially are in a highly mobile fluid, respectively semi-solid state, and the expansion proceeds significantly faster in at this initial stage than later under fully crystalline conditions. This was demonstrated Fig. 4 by the switch in expansion behaviour occurring for $t$ = 5-10 s. We believe that as a consequence, the initially faster expansion shifts the observable isopycnic shells further out to locations actually expected for much smaller densities. This explains the overall mismatch between probed and fitted densities. The stage of initial rapid expansion fastens the dilution, in particular for the densest core regions. Since the measurements in MM mode start only after complete solidification, this stage is completely missed in the experimental curves probing the largest densities. The mismatch becomes smaller at smaller probed densities. There is a well-developed maximum for $\lambda$ = 633 nm, and the effect appears to be absent for $V_{\mathrm{m}}$. The initial rapid expansion of the molten or partially solidified drop also readily explains the overall larger than expected dilution of the experimental drops.

To avoid shear melting, the extrusion process should be improved. We followed this challenge to minimize mechanical impact in additional experiments, but thus far could not obtain satisfactory results. Alternatively, future modelling could implement a time dependent effective diffusion constant during the first instances of calculation. This could be implemented without extensive numerical effort, albeit at the introduction of further free parameters and additional assumptions to be made about the crystallization kinetics far off mechanical equilibrium [55].

Second, at long times, we violate the condition of spatially homogeneous and constant particle charge and amount of screening electrolyte. This results from a slight difference in conditioning procedure for the drops and the receiving solvent. In fact, the suspension had been deionized and decarbonized, while the receiving water was merely deionized but equilibrated against ambient air. According to recent literature, dissolved $CO_2$ and its reaction products will significantly reduce the particle charge due to a combination of screening effects with pH-driven and dielectric charge regulation [35, 56]. In the present situation, $CO_2$ diffusing from the surrounding water towards the drop cores will significantly lower the particle charge in the contaminated regions and there increase $\rho_m$. After traversing the outer fluid layer, the inward migrating $CO_2$ and its dissociation product carbonic acid reach the crystalline part of the drop. In Fig. 9, this happens shortly past the maximum of $V_m(t)$. From thereon, the remaining crystalline volume shrinks much faster by $CO_2$-related melting than predicted for dilution related melting at constant deionized and decarbonized conditions. The idea of an advancing $CO_2$-front is corroborated by the very early disappearance of the blueish 221 Bragg reflections at the drop rim (cf. Video 1) and the simultaneous unaffectedness of the core volume curves. Future experiments could therefore attempt to use a gas-tight container filled with decarbonized water as receiving fluid.

An additional experimental improvement would be a buoyancy match between particles, suspending medium and receiving solvent. Under the present conditions, all drops settled to the cell bottom. There, additional influences of gravity may become operative. While for the small drops used in the present experiments, no evidence of a height dependent lattice constant was detected, there were several indirect effects observable. For instance, we observed a systematic deviation from a strictly spherical drop shape at late stages. Specifically, the presence of a well-observable particle enriched environment causes the lower drop halve to expand slower than the upper. We believe that the bottom layer of sedimented particles creates an increased osmotic pressure counteracting the expansion of the crystal sphere. To exclude an influence on our measurements, drop size analysis was solely based on data taken from the upper drop halve. To fully circumvent this effect, one may try to buoyancy-match the particles, by using low molecular weight sugar-solutions for both the drop and the environment. Preliminary efforts in that direction indeed indicate a way to prepare freely suspended drops residing in the cell centre (Fig. S6 in the SI).

Finally, we like to stress that the here employed, scattering-based microscopic imaging technique is not restricted to experiments on the expansion of crystalline drops. Rather, we anticipate that there are a number of interesting non-equilibrium experiments, which could profit from it [3, 4, 5, 55, 57]. This concerns, in particular, experiments at large particle concentrations, which so far were only accessible for hard sphere systems and a few other well-index-matched systems. Possible applications comprise a straight forward extension of the range of accessible densities in studies of crystallization, coarsening and melting kinetics, both with and without applied shear, as well as compression experiments using gravity, centrifugation or other suitable experimental techniques.

## 6 Conclusion

We developed and successfully demonstrated a novel approach to determine the density profiles in freely expanding turbid crystalline drops. We observed a promising agreement between our experiments and theoretical modelling based on dynamical density functional theory. Using highly charged spheres in an aqueous environment, we reproducibly observed an expansion scenario reminiscent, but not exactly identical to previous findings. The characteristic development should serve as a reliable starting point for systematic investigations of expansion and profiling in dependence on interaction type, shape, and strength as well as on environmental boundary conditions including their theoretical modelling. Further, our method should be readily transferable to unidirectional expansion experiments between confining surfaces or to free-standing extruded colloidal wires. We thus anticipate, that our novel approach can pave the way to density profiling also in many other situations of fundamental and/or practical interest.

**Acknowledgements**

We thank Ramsia Sreij for the X-ray characterization. We thank Daniela Kronmüller for the documentation of fully buoyant drops. We thank P. Leiderer, to whom we dedicate this paper on occasion of his 80$^{th}$ birthday, for many inspiring discussions on charged sphere phase transitions. Financial support from the DFG (Projects PA459/19 and LO418/25) and from the Inneruniversitäre Forschungsförderung der Johannes Gutenberg-Universität, Mainz, is gratefully acknowledged. G.M. Jr. was trainee within the IAESTE program and received a fellowship by the Deutsche Akademische Austauschdienst (DAAD).

**Author´s contributions**

This study was conceptualized by T. P.. Funding was acquired and administered by T.P. and H. L.. Experimental methodology was developed by J. L. M. Jr., J. A. B. W., C. H. Y, and S. J.. Analysis software was developed by E. M. and M. W.. Experiments were performed and analysed by M. W., J. R. vP.-L.; J. A. B. W., C. H. Y., S. J., and M. H.. Experiments were supervised by M. W. and T. P.. The theory part of this study was conceptualized and supervised by H. L.. Theoretical modelling and its analysis were developed and performed by G. H. P. N.. The manuscript was written by T. P., G. H. P. N., and H. L.

M. W. and G. H. P. N. contributed equally to this study.

**Conflict of Interest**

The authors declare no conflict of interest

**Data Availability**

Original data and artwork is available from the corresponding author upon reasonable request

Supplementary information is available online at: XXX

**Graphical Table of Contents:**

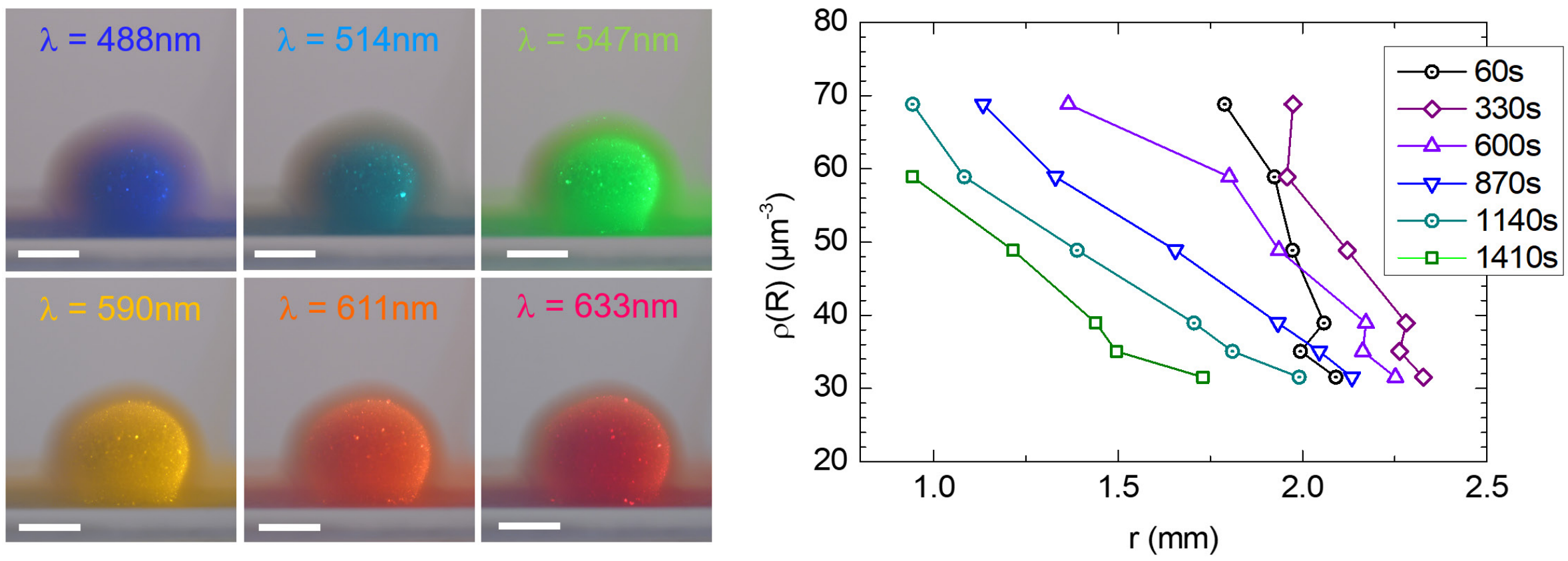

Exploiting multiple scattering, we measure the time-dependent density profile of an opalescent but turbid colloidal crystal ball in a particle-free environment.

**Supplementary material**

Accessing the Free Expansion of a Crystalline Colloidal Drop by Optical Experiments.

Marcus Witt[1], G. H. Philipp Nguyen[2], Josefine R. von Puttkamer-Luerssen[1], Can Yilderim[1], Johannes A. B. Wagner[1], Ebrahim Malek[1,3], Sabrina Juretzka[1], Jorge L. Meyrelles Jr.[1,4], Maximilan Hofmann[1], Hartmut Löwen[4], and Thomas Palberg[1]

[1]Institute of Physics, Johannes Gutenberg University, Mainz, Germany

[2]Institute of Theoretical Physics II: Soft Matter, Heinrich-Heine-Universität Düsseldorf, Germany

[3]Institute for Biophysics, Goethe University Frankfurt, Germany

[4]Pontifícia Universidade Católica, Rio de Janeiro, Brazil

**Supplementary video material**

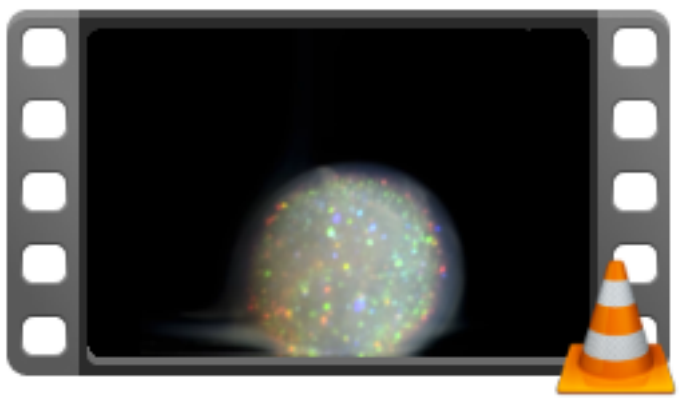

Video 1 WB.mp4

**Video 1** Expansion of a spherical drop showing vivid Bragg scattering, as observed in white-light Bragg-mode (WB) under illumination with white light from the right. Time is shown in the upper right corner. The frame size is 5.1 mm × 5.1 mm.

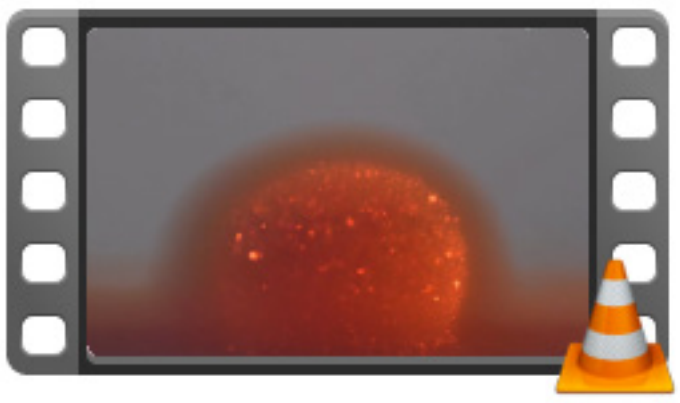

Video 2 MM 611 nm.mp4

**Video 2** Expansion of a spherical drop showing Bragg single and multiple scattering, as observed in mixed mode (MM) under illumination with monochromatic light of $\lambda$ = 611 nm from the right and diffuse white light from the back. Time is shown in the upper right corner. The frame size is 8 mm × 5 mm.

**Particle characterization**

Suspensions of exhaustively deionized and decarbonized suspensions were thoroughly characterized before use in this study [1, 2, 3, 4]. Figure S1a shows results from torsional resonance spectroscopy [2]. The shear modulus of the colloidal crystals is on the order of a few Pa. For comparison, the shear modulus of bcc iron is on the order of $10^{11}$ Pa. G increases with increasing density. The development complies well with the expectations for a polycrystalline material of bcc crystal structure. The extracted charge is related to the effective Yukawa interaction strength. In Fig. S1b, we show the linear increase in background-corrected sample conductivity with increasing density. The extracted charge is related to the effective number of freely mobile counter-ions [1, 4]. Up to 100 $\mu m^{-3}$ the sample structure could be determined by either static light scattering [5] or by reflection spectroscopy [6]. Exemplary results of the latter technique are shown in Fig. S2a. Bragg peaks are Miller-indexed for an assumed bcc structure. Note the significant increase in background scattering intensity at small wave-lengths, i.e. large $q$. It is due to incoherent multiple scattering, and its increase is in line with the expectation for Rayleigh scatterers [7]. The spectroscopic data are evaluated in Fig. S2b in a plot of $q^2$ versus $h^2+k^2+l^2$. Their arrangement on a straight line confirms the Miller-indexing of Fig. S2a. At larger densities, higher order reflections disappear in the MS-background. The positions of (200) and (211), however, keep evolving consistently with the nominally adjusted concentrations. Hence, we conclude the crystallite structure to be bcc even up to the largest investigated densities of $\rho$ = 100 $\mu m^{-3}$.

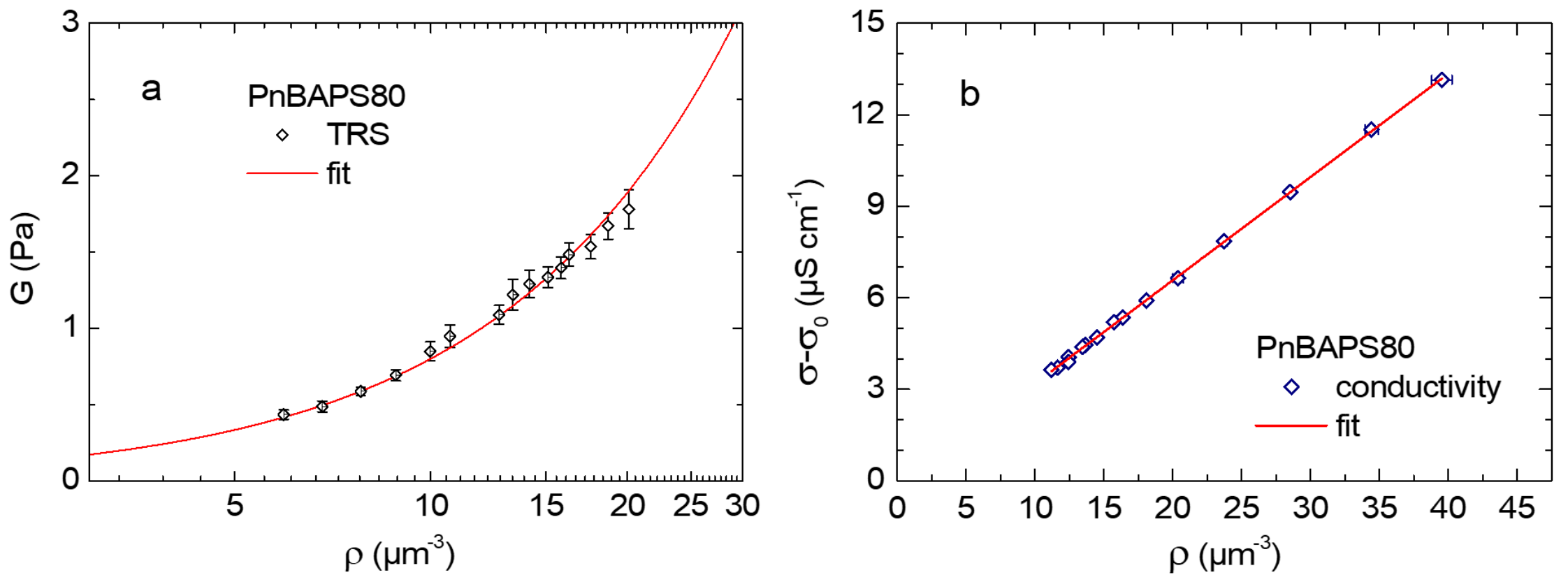

**Fig. S1 Charge characterization of deionized and decarbonized suspensions of PnBAPS80.** a) Shear modulus as determined from torsional resonance spectroscopy [2] in dependence on number density taken from static light scattering. The solid line is a two parameter least squares fit to the data with the n and the salt concentration from conductivity c = 0.2 µmolL$^{-1}$. The fit returns $Z_G$ = 365.1±2.3, where the error denotes the standard error at a confidence level of 0.95. b) Background-corrected conductivity in dependence on number density. The solid line is a two parameter least squares fit to the data using Hessinger's model of independent ion migration [1]. The fit returns Z = 513±3, where the error denotes the standard error at a confidence level of 0.95.

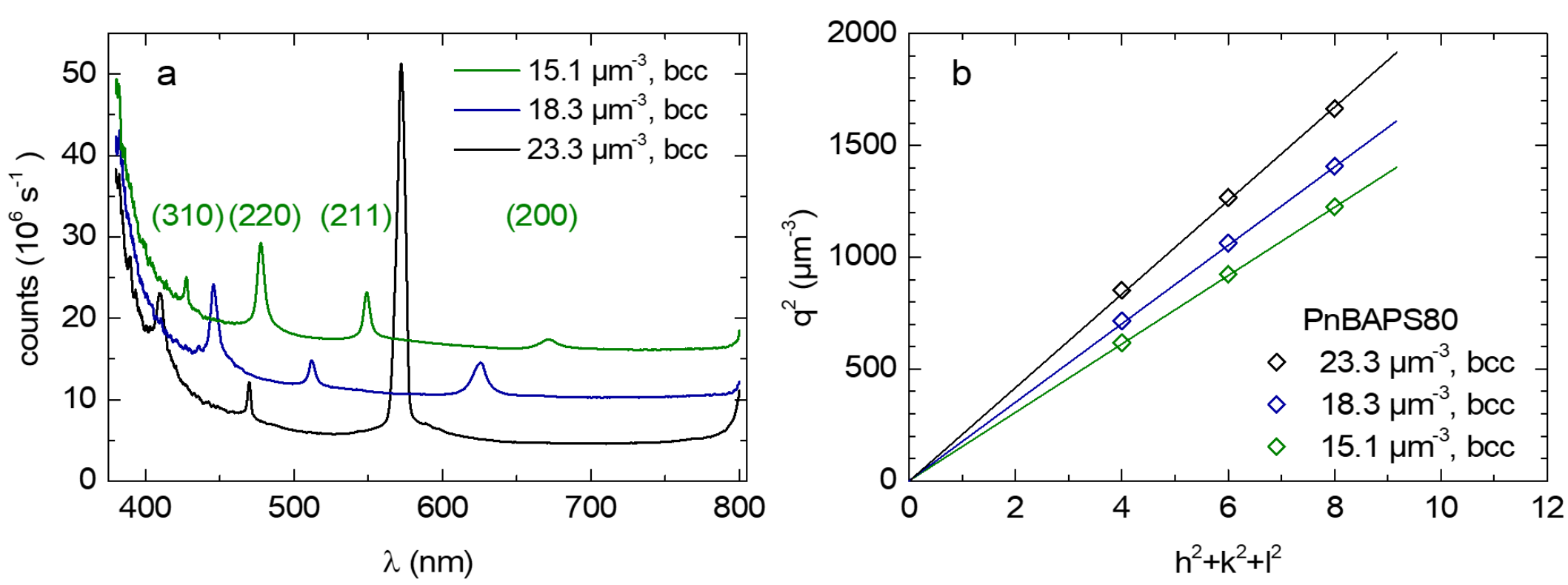

**Fig. S2 Structure and density determination on deionized PnBAPS80.** a) Miller indexed Debye-Scherrer scattering pattern from reflection spectroscopy for three different number densities indicated in the key. Curves are shifted for clarity. b) Plot of $q^2$ versus $h^2+k^2+l^2$. The solid lines are two-parameter least squares fits to the data using Bragg's formula and indexing the observed peaks for bcc crystal structure. From the slope, the fits return the ρ-values indicated in the key. The relative standard error at a confidence level of 0.95 is about 1.2%. At densities larger than 45 µm$^{-3}$, the scattering patterns displayed only two or even only one peak. This raised the relative error in density determination to about 2% and 4%, respectively.

For a known interaction effective charge, deionized conditions and increasing density, one may calculate the so-called state-lines [8] and compare them to the predictions of the phase behaviour of Yukawa particles. Following [9], it is convenient to do so in the plane of the effective temperature ($T_{\mathrm{eff}} = k_B T/V(R)$ and the coupling parameter $\kappa R$ as shown in Fig. S3. In this plane, the sample will be fluid as long as its state point is above the melting line. Various authors have calculated the latter [10]. We here compare our state line to the results of Robbins et al. and of Meijer and Frenkel [9, 11]. The crystalline regime to the lower right shows in addition the location of the bcc-fcc transition expected to occur at elevated densities. The state lines reveal a shallow maximum under deionized conditions and a monotonous decrease of the effective temperature with increasing density at an underlying salt concentration of 6.3 µmolL$^{-1}$. The experimentally probed densities are shown as small coloured symbols. We emphasize, that our experimental melting point agrees well with the predictions. The starting density is just inside the theoretically expected fcc region, but it is known from many other samples, that the bcc-fcc bulk transition is most often suppressed by kinetic effects such that only bcc is observed even at large κR [12]. Hence, we anticipate that we start in a bcc state. Note that during expansion, κ decreases by a factor of about 2 while $T_{\mathrm{eff}}$ decreases only very weakly.

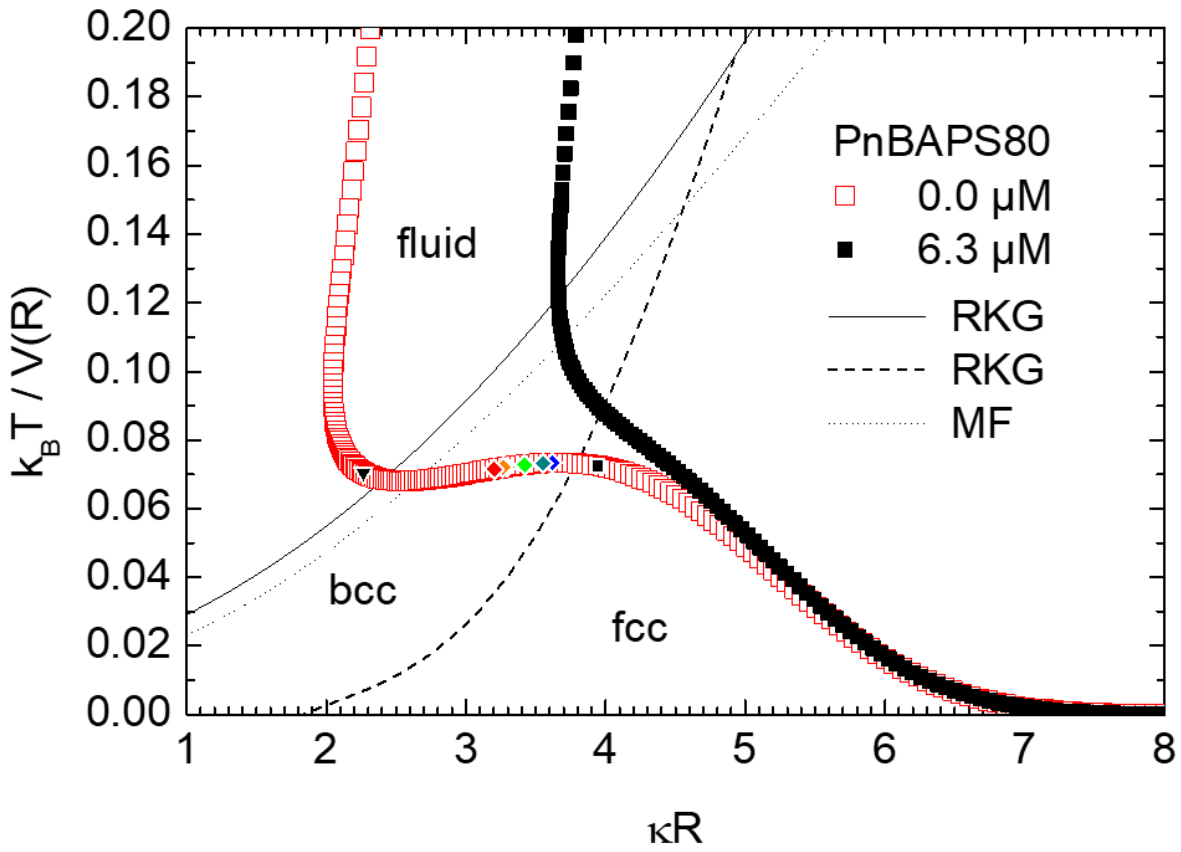

**Fig. S3 State diagram and state lines of the PnBaPS80 particles in the effective temperature – coupling parameter plane**. Data for the deionized and decarbonized state (red open diamonds) are compared to data at an electrolyte concentration of 6.3 µmolL$^{-1}$. The small symbols denote the state points for the experimentally probed densities in the expanding crystalline droplets. The black triangle denotes the melting density, the black square denotes the start density. The coloured diamonds denote the densities as probed by monochromatic Bragg scattering at different wave lengths of 488 nm, 514 nm, 547 nm, 590 nm, 611 nm and 633 nm, corresponding to densities of 68.83 µm$^{-3}$, 58.91 µm$^{-3}$, 48.87 µm$^{-3}$, 38.95 µm$^{-3}$, and 31.54 µm$^{-3}$, respectively. The solid and dashed lines give the location of the melting line and the bcc-fcc transition, respectively, as predicted from simulations by Robbins et al. (RKG [9]); the dotted line shows the predictions of Meijer and Frenkel (MF [11]).

**Additional details concerning the observation optics**

Figure S4 shows an extended drawing of the optical arrangement used in the imaging experiments in top view with annotated comments. All elements are mounted on optical rails collinear with the optical axes (not shown), such that a facile rotation of the respective illumination and observation directions by 90° becomes possible. This turned out to be very useful for drop shape inspection.

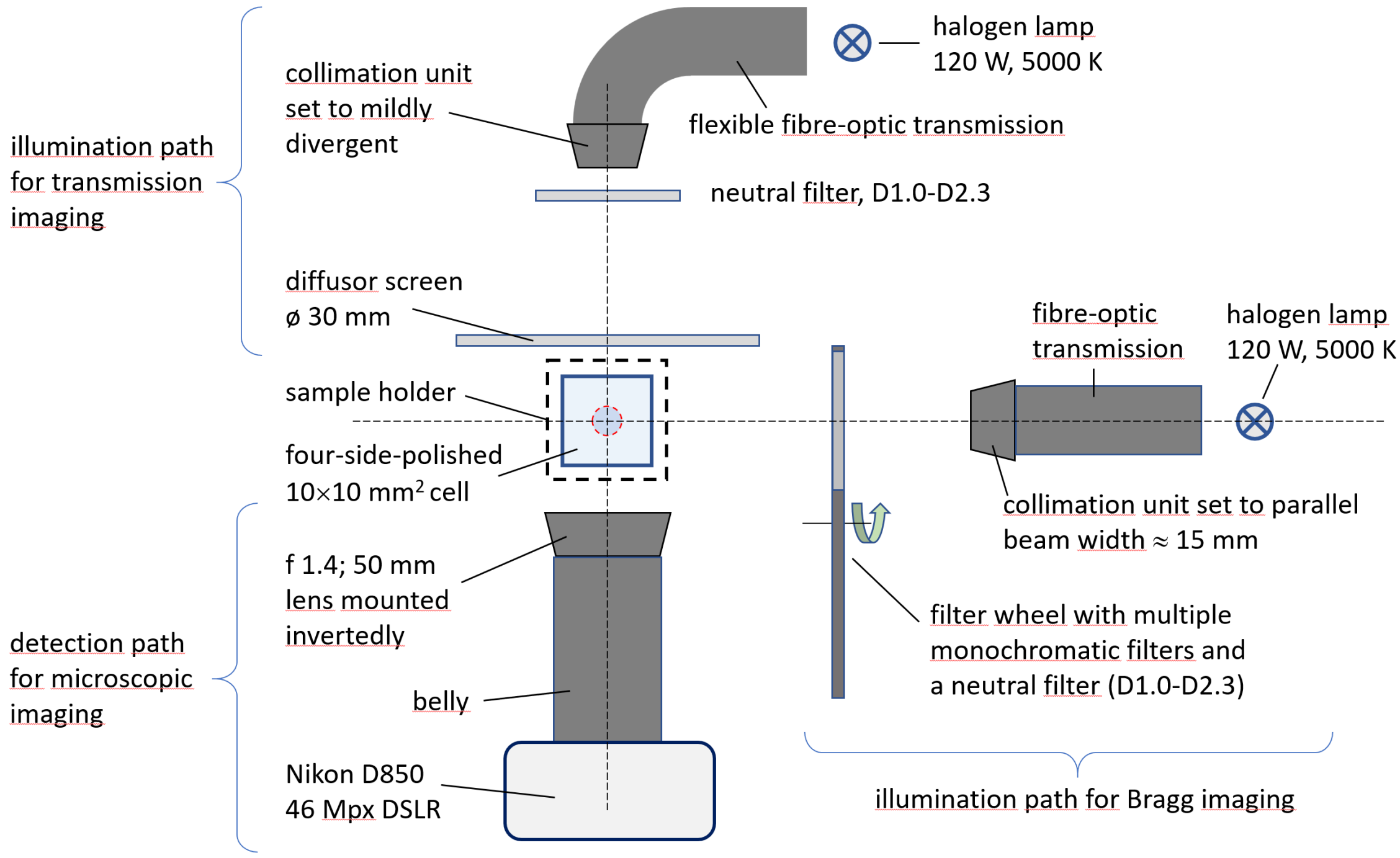

**Fig. S4 more detailed optical set-up in top view.** All items are labelled in this drawing. Commercial optical components used in illumination and observation are shown in dark grey. Additional optical elements for illumination adjustment are sketched in light grey. The vertical and the horizontal dashed lines indicate the optical axes of illumination and observation paths, which are oriented either at right angle to each other or collinearly.

**Drop shapes**

Drop compactness and shape are crucially influenced by the way the drops reach the observation position. Tip position with respect to the meniscus centre, fall height and the extrusion velocity crucially combine to determine the drop fate. Placing the drops off-centre with respect to the cell meniscus may lead to strong anisotropic distortions. Much too large fall heights lead to drop splitting, slightly larger than optimum fall height or fast extrusion promote doughnut-shapes. Too small fall heights result in drop adhesion to the top water surface. Some examples are shown in Fig. S4. With some practice, however, we regularly obtained compact, spheroidal or ellipsoidal drops. The following standard protocol was adapted for working in a 1cmx1cm cross-section cell: DI meniscus height 15mm, fall height 5-7mm as measured from the flat syringe needle tip to the water surface, slow extrusion and drop self-release by gravity. Quite reproducibly, this resulted in compact drops of prolate or oblate ellipsoidal shape, many of these being nearly spherical (Fig. S6b). We explicitly note, however, that irrespective of shape, the general free expansion and melting behaviour was well reproducible in all cases.

To further judge the shape, ellipsoidal candidates were inspected in either transmission or scattering also from the left direction in Fig. 1b, i.e. from 90° off the standard observation direction. Drops qualified for further evaluation when, upon inspection from both sides, the first eccentricity was below 0.14. This quantity was calculated using $\varepsilon = \sqrt{1 + a^2 / c^2}$ for the ellipsoids derived from fitting the projection of the MS-scattering surface by ellipses. In the examples of Fig. S6b the two rightmost drops qualified for further evaluation.

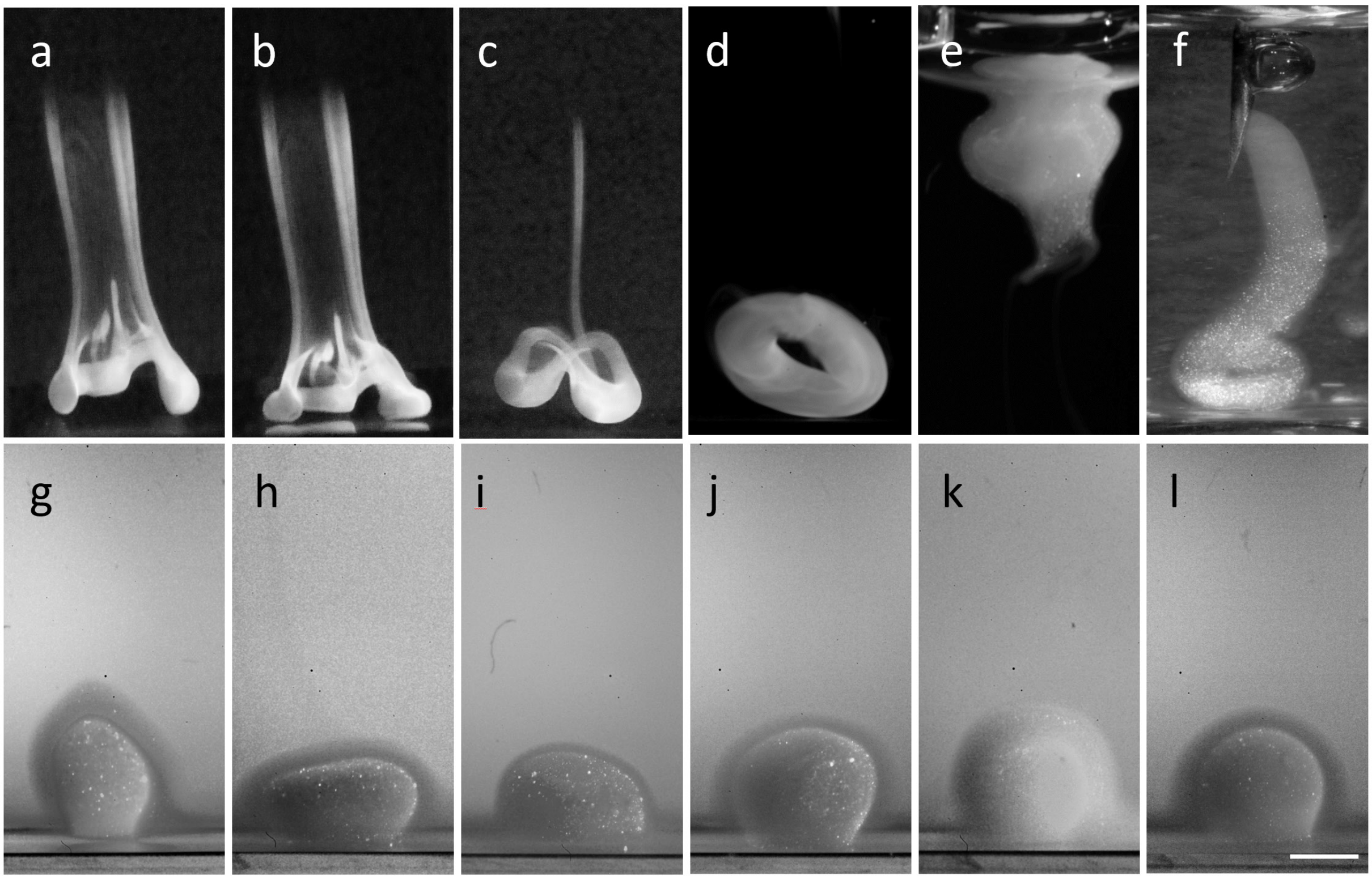

**Fig. S5 Variation of drop shape.** a) examples of drop shapes obtained under non-standard conditions. B/W renderings of WB images taken at very early times. a) and b) Much too large fall heights leads to drop splitting. Images were taken 1s apart just before and after touchdown. Note the easy deformation upon touchdown. c) and d) Too Large fall height leads to doughnut-shaped drops, which become more regular at decreasing fall height. e) Too small fall heights often result in substantial drop fractions adhering to the Water-air interface. f) extrusion with the needle tip inside the water produces wormlike shapes. g) to l) Drops produced by the standard extrusion procedure regularly show compact spheroidal shapes. B/W renderings of MM images taken at t ≈ 300 s. The scale bar is 2 mm. All drops appear to have a flat bottom. Shape determination was therefore performed on their upper hemisphere, not influenced by gravity and/or adjacent fluid sediment.

Very recently, we started experiments aiming at buoyancy-matched drops. To this end, the suspension was prepared from sugar solutions (analytical grade glucose, Merck, Germany) instead of DI-water. We checked, that subsequent deionization did not alter the suspension composition nor the particle properties. The final suspension had a mass density of $\rho_{mass}$ = 1.005 gcm$^{-3}$. The same sugar concentration and density was adjusted in the receiving particle free solution. Experiments repeated with these ingredients regularly show buoyantly hovering, freely suspended drops, which as before expand and shrink. Drops appear more compact than before with significantly less suspension sheared off upon impact. However, most drop shapes were toroidal or mushroom-shaped. Two typical examples are shown in Fig. S5. Experiments aiming at spherical drops continue.

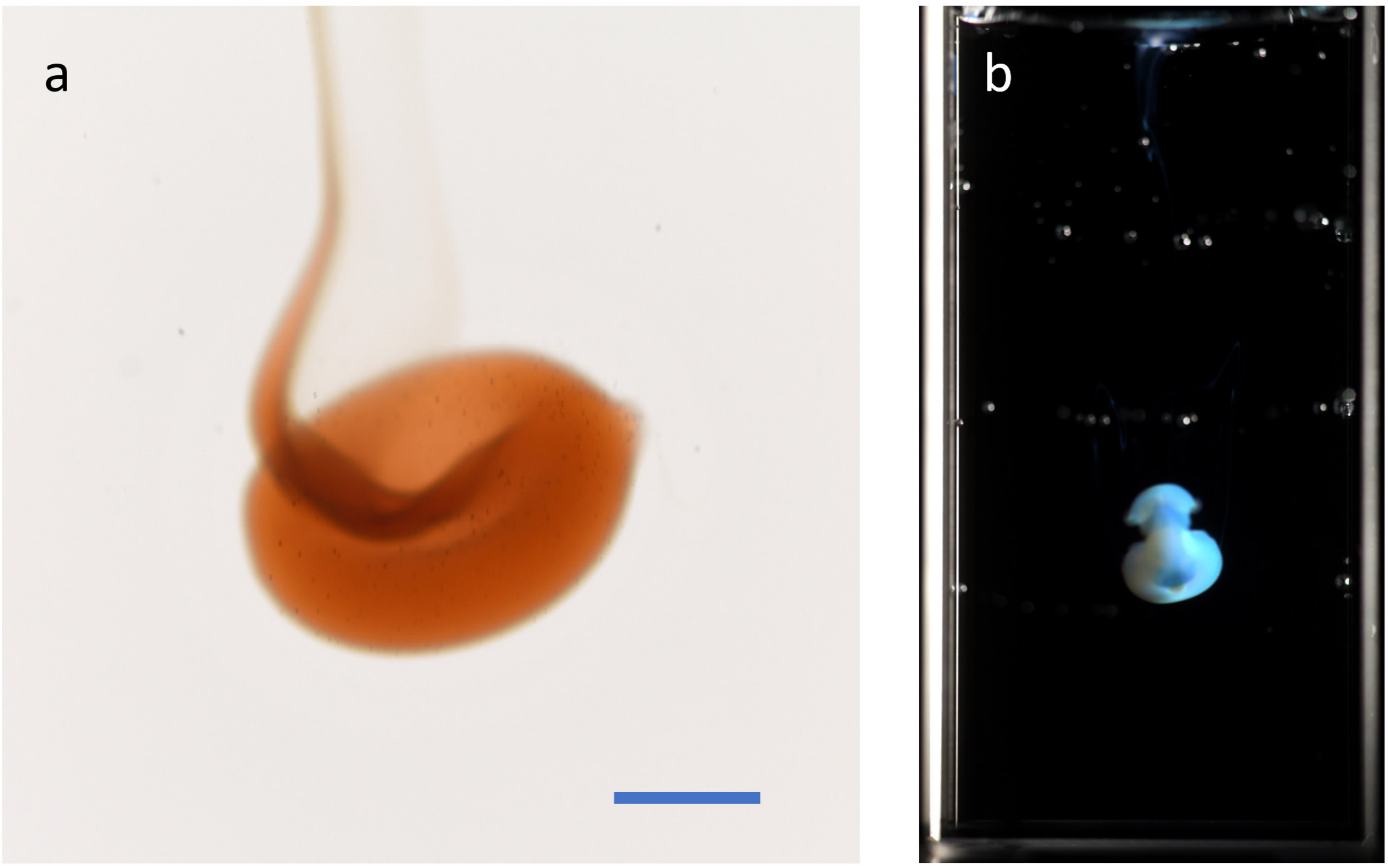

**Fig. S6 Preliminary result from buoyancy-matching experiments.** a) TR-image of a buoyant toroidal drop taken 20s after dropping. The scale bar is 2mm. b) WB-image taken 3 min after release of this freely suspended drop. Bath dimensions 10mm × 10 mm × 25mm.

**Characterization of sample scattering properties**

We characterized the scattering properties in separate experiments using transmission and reflection spectroscopy in slab geometry. The results of reflectometric structure analysis and number density calibration have been shown above. Since particles are non-absorbing, all extinction relies on scattering. Light is lost by both coherently and incoherently scattered light. Within Born approximation (single scattering), the intensity of *incoherently* scattered (IS) light is $I_{inc} \propto I_{f,0}\ n\ b(0)^2\ P(q)$. Here, the prefactor $I_{f,0}$ comprises the experimental details. It depends on the wavelength as $\lambda^{-4}$ and on the distance between drop and sensor as $R_D^2$. $b(0)^2$ denotes the low $q$ limit of the single particle scattering cross-section, $n$ is the particle number density, $P(q) = b(q)^2/b(0)^2$ is the form factor. Incoherent scattering arises from size polydispersity [13, 14] and for the present small particles of $PI = 0.08$ a large incoherent scattering cross-section $b(0)^2\ P(q)$ is obtained at any angle. The *coherently* scattered intensity is $I_{coh} \propto I_{f,0}\ n\ b(0)^2\ P(q)\ S(q)$, where $S(q)$ is the structure factor [15] which depends on crystal structure and orientation. Figure S6a sketches the experimental scattering situation for bcc (110) with the illumination coming from the right and the outgoing beam impinging on the detector. In the present case of high densities, we have finely grained, polycrystalline samples showing a Debye-Scherrer scattering pattern in static light scattering and a multicoloured opalescence in WB imaging.

Bcc crystals are stable for $\rho \geq \rho_m = 15\ \mu m^{-3}$. In the drop expansion experiments, they are observed at wavelengths of 488 nm $\leq \lambda \leq$ 633 nm and at $\Theta = 90°$. Depending on density, different reflections contribute to recorded scattering (cf. Fig. S2a). The scattering vector of any Bragg reflection equals the reciprocal of a lattice plane spacing, $q = 2\pi/d_{hkl}$. For bcc crystals, the sum of Miller indices $h,k,l$ is even and in the probed observation range we may only expect (110), (200) and (211) reflections. The first order reflection occurs at the MS-core surface. It is seen in MM images as a uniformly coloured sphere (Fig. S6b), The second order appears in the enveloping region. The third order is visible only at the crystal sphere rim and shifts out of range for $n \geq 18.2\ \mu m^{-3}$ (Fig. S6c). By contrast, in transmission, all reflections in the range $0° < \Theta < 180°$ contribute to extinction.

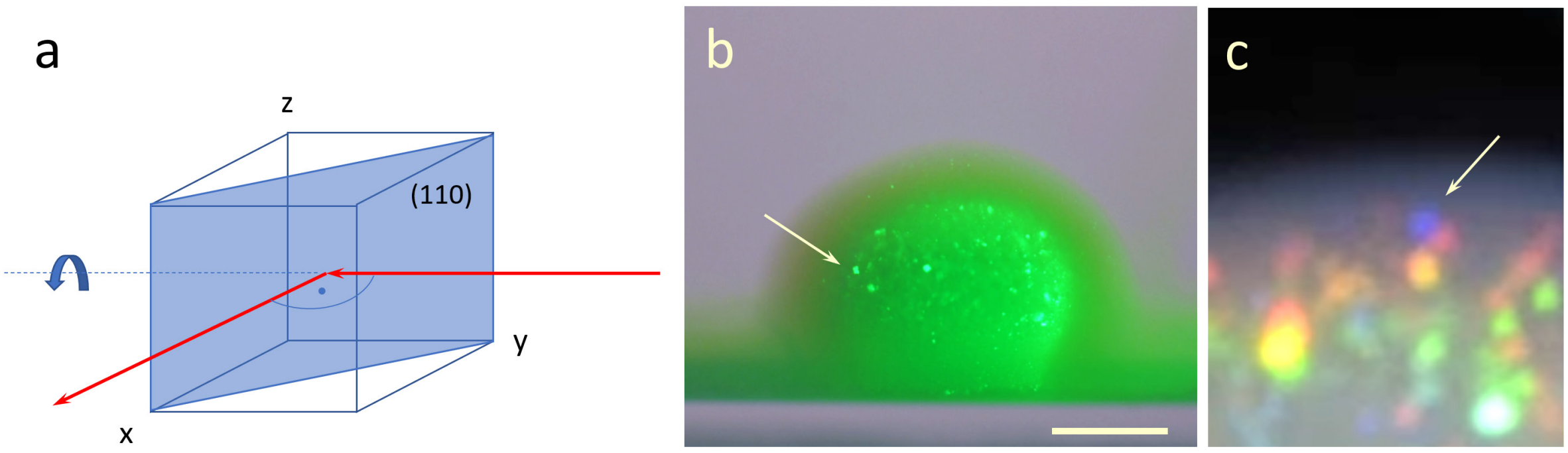

**Fig. S7 Imaging details.** a) Schematic drawing of a bcc crystal, suitably oriented of for (110) scattering at 90° towards the detector. Note that $\vec{k}_f$ lies within the locally flat isopycnic plane. By turning the crystal about the illumination optical axis, it can scatter in any direction within this plane and illuminate further material. b) MM observation of the multiply scattering (110) Bragg shell at $\lambda$ = 547 nm corresponding to $\rho_C = 48.88\ \mu m^{-3}$. The arrow highlights a reflection occurring on the backside of the drop, which is not illuminated directly. The scale bar is 2 mm. c) Cropped WB image showing a blueish (211) Bragg reflection marked by the arrow. Only very occasionally, and only at rather early times, such 3rd order Bragg reflections are observed at the very rim of the crystal phase.

With increasing density, Born's approximation is invalidated, and we encounter weak to strong multiple scattering (MS) for both coherently and incoherently scattered light. MS has been studied in samples of homogeneous density and is generally considered detrimental for static and dynamic light scattering as well as for image analysis. Much effort has therefore been devoted to the isolation of the single scattering signal [16], but only few studies actually exploit its properties by sophisticated instrumentation. To characterize the general scattering and transmission characteristics for later use in the analysis of our main experiment, we performed white light transmission (TR) experiments in slab geometry. We placed a layer of concentrated suspension at the bottom of a rectangular cell of cross-section 5 mm $\times$ 2 mm and left it to expand uniaxially against the supernatant DI-water. During expansion, we recorded the transmitted light with a DSLR consumer camera. The deposited suspension expanded very slowly. Uniaxial expansion in slab geometry corresponds in principle to the free drop expansion studied in the main experiments. However, the influence of gravity and sample wall interactions severely influences the expansion dynamics in slab geometry and the resulting expansion dynamics should not be directly compared to the free expansion experiments.

A key feature of the slab geometry is that projection effects are avoided. We could access the wavelength dependent transmission of light in a more dilute sample as well as the relative contribution of different scattering mechanisms in different regions in a concentrated one. In Figure S7a, we show transmission profiles obtained at t = 1400 s in a fluid ordered sample with a starting density of $\rho = 20\ \mu m^{-3}$. We compare total transmission to the individual readings of the RGB channels. The observed dependence on wavelength demonstrates the expected $\lambda^{-4}$ dependence of the prefactor $I_{f,0}$. Note, however, that due to the restricted wavelength resolution of the RGB channels the profiles obtained under white-light illumination are not strictly single exponential as would be expected from a Beer Lambert law and as indeed observed under monochromatic illumination [17, 18].

In Fig. S7b, we studied a sample with starting density $\rho_0 = 110\ \mu m^{-3}$, which evolved layers of different structure and optical properties. We here set the illumination to maximum intensity. This leads to sensor saturation already at weak to moderate extinction (region labelled ⑤) but yields a detailed profile at larger extinction. The arrow marks the melting transition, located between regions ④ (fluid) and ③ (crystalline). Note the change in slope of the extinction curve. Above the crystalline/fluid transition, both scattering types contribute. However, with rapidly radially decreasing ρ and the simultaneous loss of fluid structure, a radial dependence of the transmitted light results which is more pronounced than in the crystalline region. Below the transition, extinction is mainly due to incoherent scattering and only little due to individual crystallites favourably oriented for (200) Bragg scattering. Thus, we observe a change in slope. Note further the small z-extension of the (200) Bragg scattering region ③ in these uniaxial expansion experiments, presumably due to gravitational compression by the fluid layer on top. Just below this region, we observe a deep transmission minimum denoted ②. In visual inspection of samples illuminated from the right by white light, its height correlates well with that of a narrow but vividly scattering horizontal feature showing an RGB colour-banding with red at its top. We attribute the transmission minimum to coherent MS due to strong 110 Bragg scattering occurring in layers of suitable density.

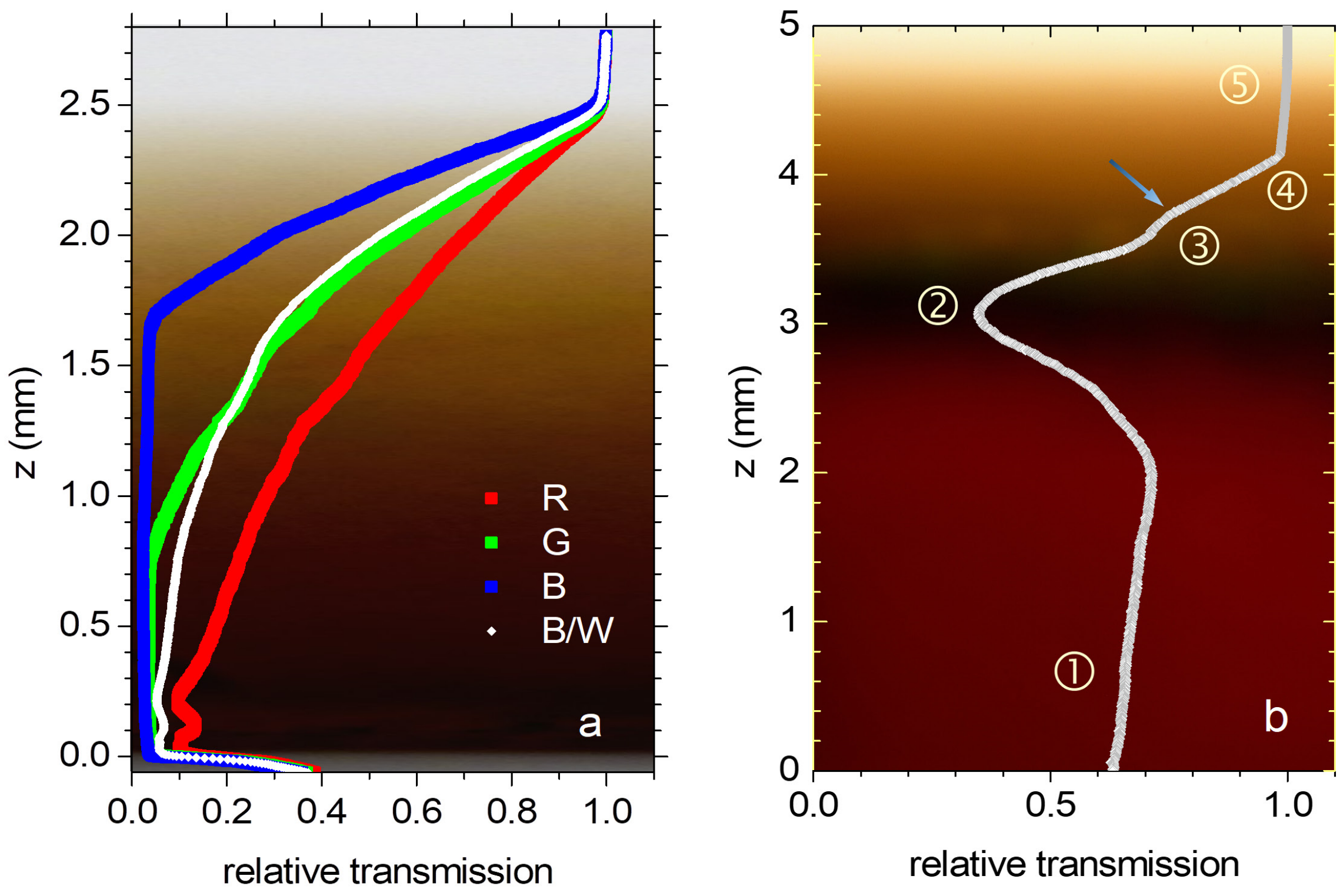

**Fig. S8 Transmission experiments in slab geometry.** In both panels, the background shows the original TR images. a) Relative transmission profiles obtained at t = 1400s in a fluid ordered sample with a starting density of $\rho_0 = 20\ \mu m^{-3}$. We compare total transmission to the individual readings of the RGB channels. Note the pronounced increase of extinction with decreasing wavelength. b) Transmittance profile as a function of height above ground. The numbers mark regions with distinct characteristic developments of the relative transmission with height above ground. ①: moderately transparent bottom region with linearly rising transmittance, ②: MS extinction band; ③: continuation of the moderate extinction at a slightly increased level due to additional 2nd order Bragg scattering crystals; Note the sheet of crystallites in this narrow region faintly visible in the background image; ④: steeply decreasing extinction in the fluid region. The border between ③ and ④ is marked by the grey arrow. The flattening of the transmittance curve beyond Z = 4 mm (region ⑤) is an artifact due to sensor saturation. For further details, see text.

Note that this coherent mechanism is self-restricting in the sense, that photons may propagate horizontally within the layer under continuing coherent scattering. Photons scattered into other regions, where the Bragg condition is not fulfilled any more, may there only scatter incoherently. In the drop expansion experiments, MS light propagation occurs in an analogous way within the spherical isopycnic shell. This can be illustrated using Fig. S6a. If the sketched crystal is rotated about the optical axis of the incoming beam, the resulting 90° Debye-Scherrer cone lies exactly within the locally flat isopycnic plane. In fact, Bragg scattered light can propagate throughout this plane and give rise to further reflections at suitably oriented crystallites. This wavelength- and density-dependent coherent MS is enhanced in particular at early stages of drop expansion, when crystallite sizes are still small, and the Bragg reflections show some Scherrer broadening [19]. This enables secondary scattering events also at less accurately oriented crystals and facilitates MS-light propagation within the whole of the shell up to the occurrence of Bragg reflections also within the core-shade. This is highlighted in Fig. S6b by the arrow in the shown MM-image recorded under illumination with light of $\lambda$ = 548 nm. In WB images, this light propagation leads to the occurrence of concentric coloured bands and regions of near equal brightness on the right and left side of the drop centre (cf. Fig. 2e and Video 1). Overall, MS thus enhances imaging contrast and allows crystal sphere sizing as well as density profiling in the drop interior.

Coming back to the extinction experiments in Fig. 5b, the locally increased photon density within the MS-layer will, however, also result in a significant increase of incoherent scattering from within the layer and in its vicinity. Thus, in extinction, the width of the transmission minimum appears enhanced as compared to direct visual inspection. This effect is well-known as blurring in optical microscopy on turbid samples [41], and in the main scattering experiment, it affects spatial resolution in particular at short wavelengths. Finally, in the bottom region ①, the crystal lattice constant is too small to give rise to any Bragg scattering. Consequently, this region is again moderately transparent. Extinction occurs solely due to incoherent MS, and the transmission decreases nearly linearly with decreasing z.

Summarizing: the white light transmission experiments clearly demonstrate that a) 110 Bragg scattering causes pronounced extinction confined to a narrow band of specific density, which will narrow further under single wavelength illumination; b) a characteristic change of slope is associated with the melting transition; and c) throughout the crystalline core region we have only moderate extinction by incoherently light. These findings imply that: i) the core region of the drop is inaccessible to WB and MM imaging; ii) in TR images, we can determine the location of the melting transition from the change in radial slope of transmittance; iii) in MM images, the vivid 110 scattering seen for different wavelengths is in fact originating from thin ellipsoidal shells of constant density. Monitoring their wavelength- and time-dependent location will then allow for time-dependent density profiling.

### Additional series of processed WB and MM images

For illustration purposes, Fig. S8 shows three additional time series taken in WB and MM mode. In these MM images, the drop is monochromatically illuminated from the right and simultaneously from the back. Note the excellent visibility of (200) reflections in the broadening, more transparent outer regions of the crystal sphere in Figs. S8 c-h. Note further the gradual coarsening of the core structure. This is also seen clearly in Fig. S8 i-q and r-y. In both modes, we observe (110)-scattering also occurring from the "backside" of the sphere, which is not directly illuminated (e.g. Fig S8e, f, k and u). This nicely illustrates MS light propagation within the respective (110) shells. The images in the two MM series were recorded directly after one another. Note that some (but not all) individual reflections are visible in both corresponding colour panels. This indicates a radial density inhomogeneity within individual crystals.

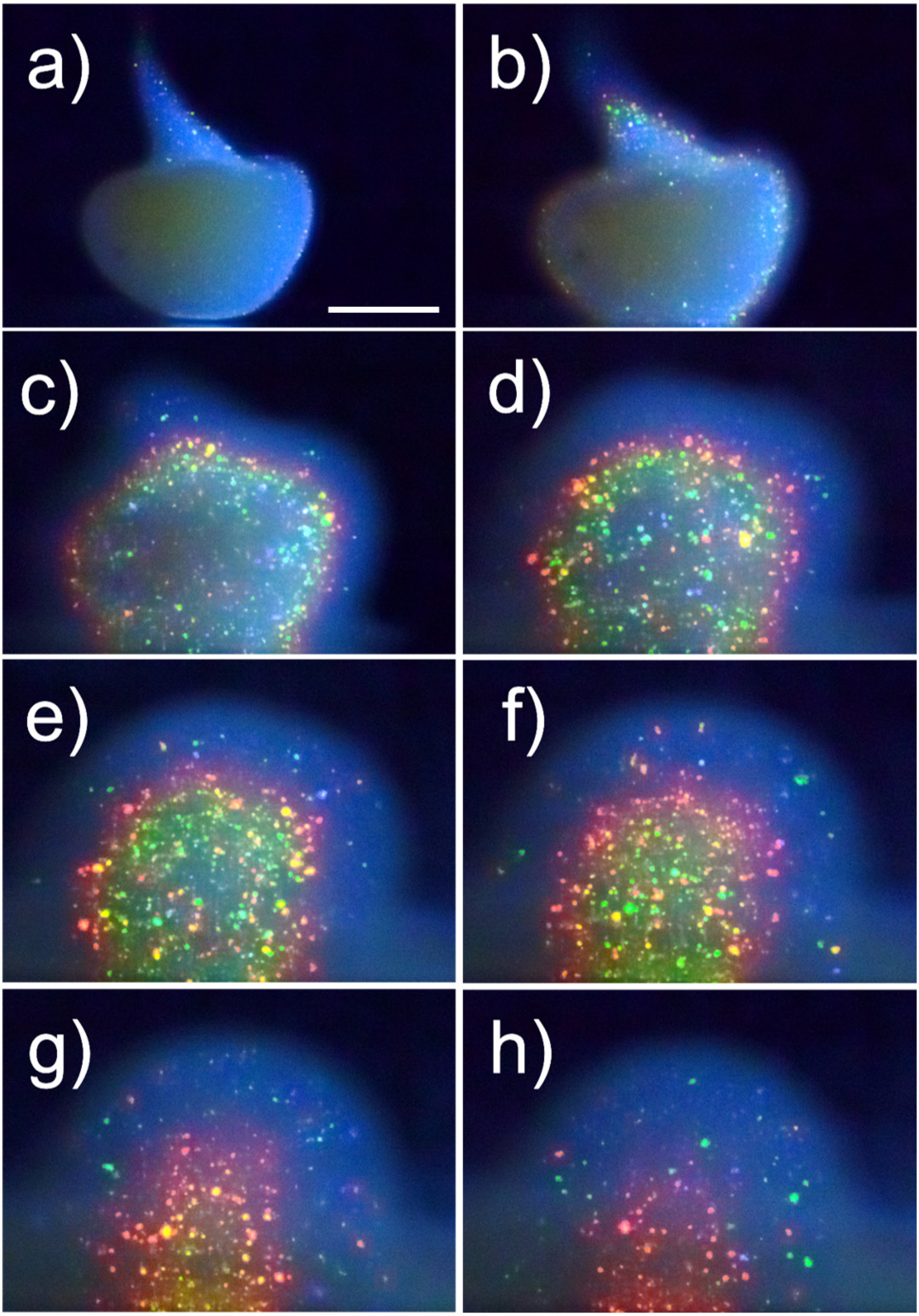
a)
b)
c)
d)
e)
f)
g)
h)

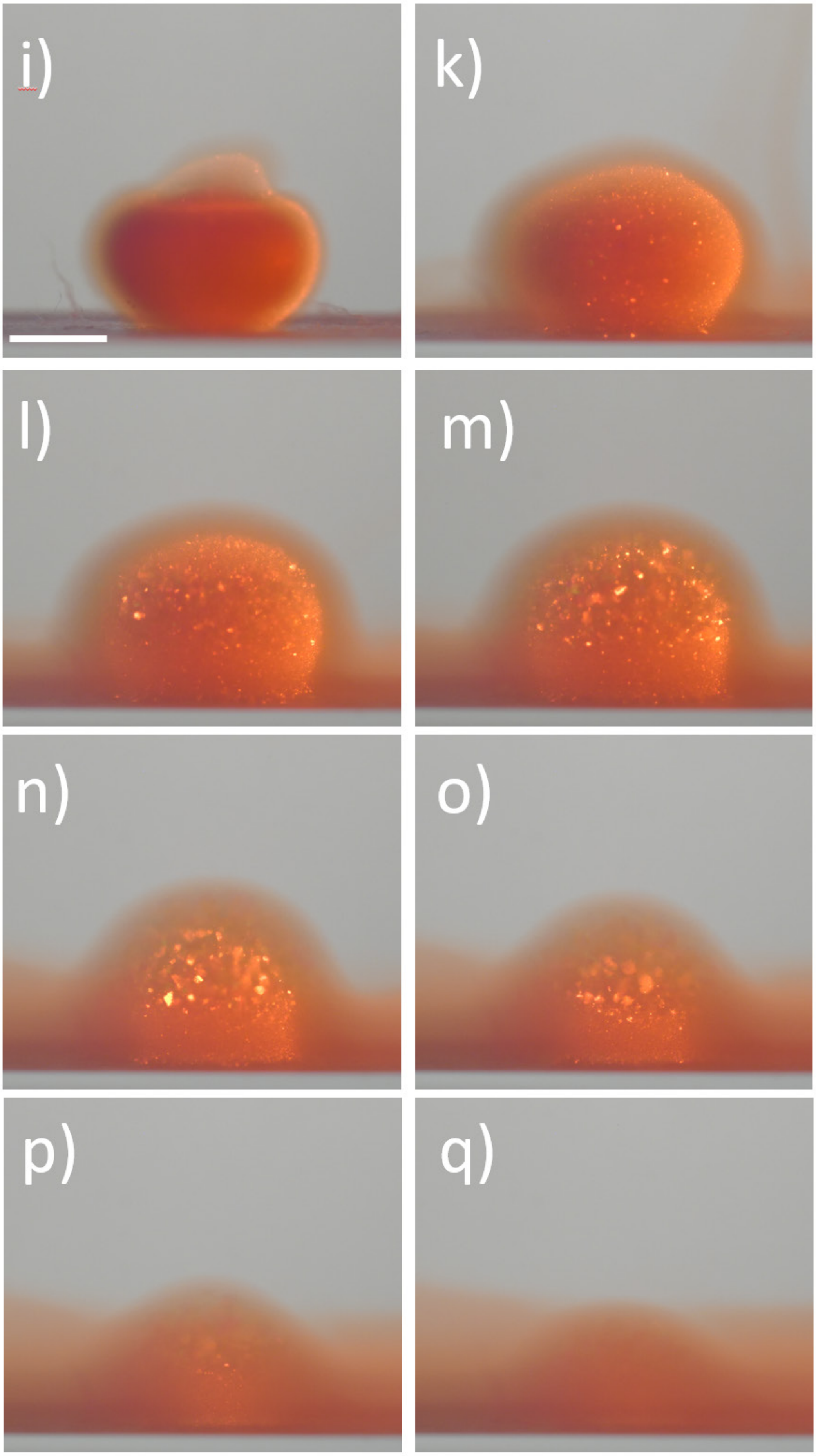
i)
k)
l)
m)
n)
o)
p)
q)

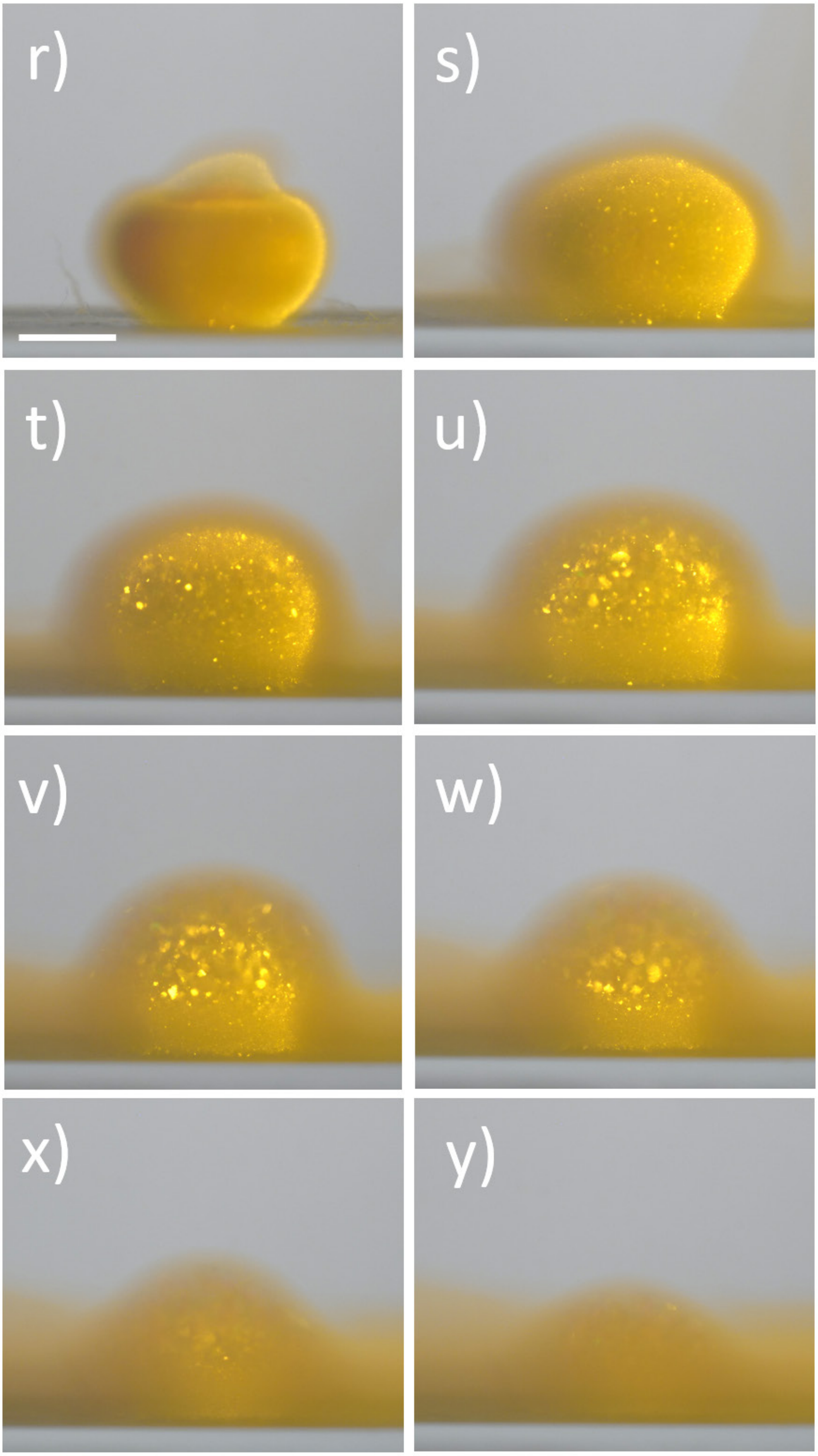
r)
s)
t)
u)
v)
w)
x)
y)

**Fig. S9 Example time series for two drops observed in different modes.** The Scale bars are 2mm. a)-h) Series of processed WB images of the drop shown in Fig. 2e. During processing, images were centred, aligned, white balance corrected for a colour temperature of 5000 K, cropped, and size calibrated. The scale bar is 2 mm. i)-q) MM image series for $\lambda$ = 611 nm. Images were recorded at t = $\{$60 s, 360 s, 660 s, 960 s, 1260 s, 1560 s, 1860 s, 2160 s$\}$ and processed after capture (centred, aligned, white balance corrected using a 5x5 pixel area of the background, cropped, and size calibrated). r)-y) The same but for $\lambda$ = 590 nm.

**Extracting wavelengths from WB images for density profiling**

As mentioned in the main text, we attempted in WB imaging to extract the scattered wavelengths of (200) Bragg reflections from the corresponding RGB readings using a recently proposed approach [20]. It corrects for the hue dependence of the RGB readings by normalizing each channel to the total intensity, and it originally was applied to the analysis of changes in the dominant colour of Gecko skin. Applied to expanding crystals showing (200) scattering, however, it failed to produce reliable results. The main difference between the two experimental situations is the use of monochromatic sensor illumination by the Bragg reflections, which leads to a positioning of the provided sensor illumination outside the RGB triangle in the C.I.E. 1931 chromaticity diagram [21, 22]. We therefore tested our sensor for its reaction to this type of illumination. The light of a halogen lamp (Avalight-DH-S; LS-0610025, Avantes B.V., Apeldoorn, NL) was collimated, dimmed by a variably neutral filter and diffracted using a line grating. From the continuous 1$^{st}$ order diffraction pattern, a tiny spectral range ($\Delta\lambda \leq$ 0.8nm) was selected by a slit aperture and allowed to impinge on the CMOS sensor of the camera. The RGB readings are compared in Fig. S9a to the independently recorded spectrum of the lamp (AvaSpec-2048-SPU2-FC 286 nm-841 nm; Avantes B.V., Apeldoorn, NL). Under quasi single-wavelength illumination, significant secondary red and green maxima are observed in the sensor RGB readings short-side the main red and green maxima (arrows). These effects became more pronounced at increasing intensities. Applying the procedure suggested by [20] then yields a near linear but not strictly monotonous relation between observed and assigned wavelength. This is shown in Fig. S9b. In particular, the two regions between 520 nm and 570 nm and below 470nm are strongly affected, and no unequivocal wavelength identification is possible there. The tested approach thus renders assigned densities unreliable. To exclude an influence of the specific high resolution camera sensor of the D850, we also tested several other Nikon SLRs including the one used in [20](D700, D750, D800, and D810) consistently reproducing the effect of secondary maxima. We further analysed white light passing a diffusor screen instead of a grating. Quite remarkably, no pronounced secondary maxima in the RGB sensitivity were observed. Rather, this time the spectral shapes featured single maxima with monotonous rise and decay. Moreover, this behaviour was independent of intensity. We therefore suspect that the present failure of the approach in interpreting MM-images is related to the use of monochromatic light entering the camera and leading to issues with the hue calibration. In hindsight this might be less surprising, since the tested procedure was designed to work for illumination by light lying inside the RGB triangle, and not from the borders of the C.I.E. colour-space.

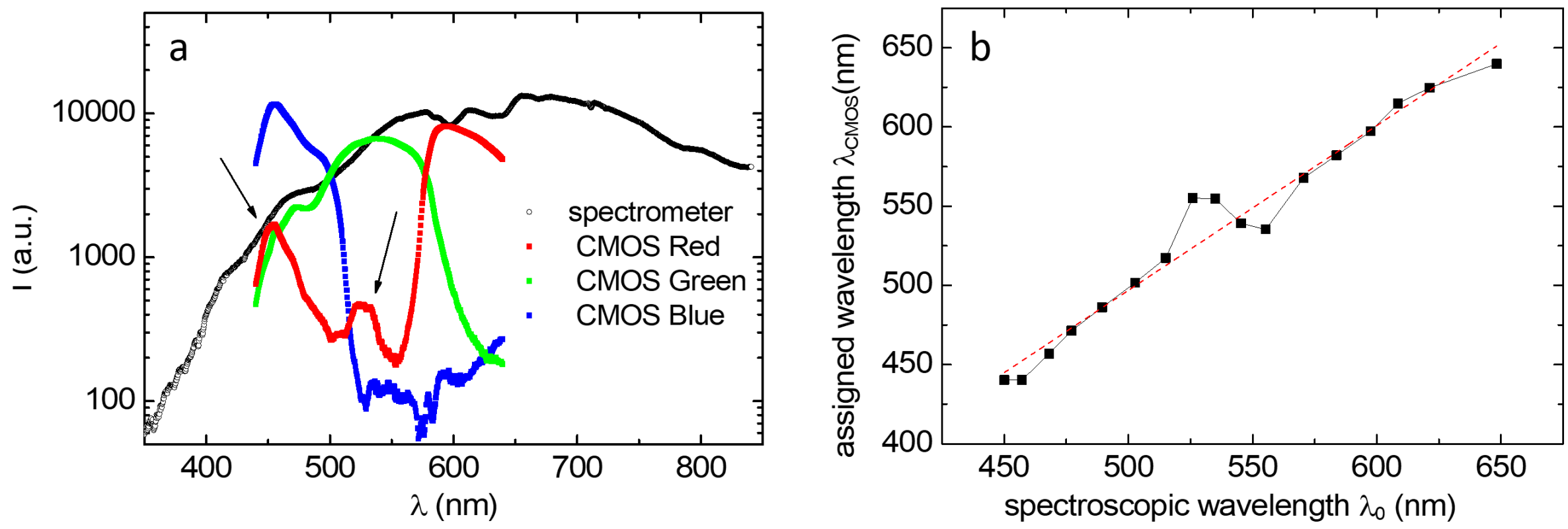

**Fig. S10 Assignment of illuminating wavelengths to readout RGB values.** a) Spectrum of the halogen lamp used for illumination in the experiments (black) compared to the corresponding CMOS readouts for the three colour channels R, G, and B, as recorded under single wavelength illumination. The arrows mark secondary maxima. b) Calibration curve resulting from evaluation of RGB readings with the procedure suggested by Teyssier et al. [20].

**Additional information concerning model calculations**

We studied the evolution of the central density in our DDFT calculations. We used the following (reduced) standard parameters to obtain and display the data shown in Fig. S10:

$d$ = 80 nm

$D = 4D_0$ with $D_0 = k_BT/3\pi\eta d$

$Z$ = 365

$\rho_0$ = 110 µm$^{-3}$

$V_0$ = 11.5 mm$^3$

$\eta$ = 1 $10^{-3}$ Pas

$a_0 = (2/\rho_0)^{1/3}$

$r_0 = 10^3 a_0$

$\kappa = 3a_0^{-1}$

$U = 10^3 k_BTa_0$.

The reduced time unit is $t = 10^3\, a_0^2D$.

For very short times, the central density remains constant (see inset). It then decays in a slowing fashion. The latter is due to the gradual vanishing of the density gradient in the core centre. At long times, the drop core region thus shows a homogeneous and isotropic expansion.

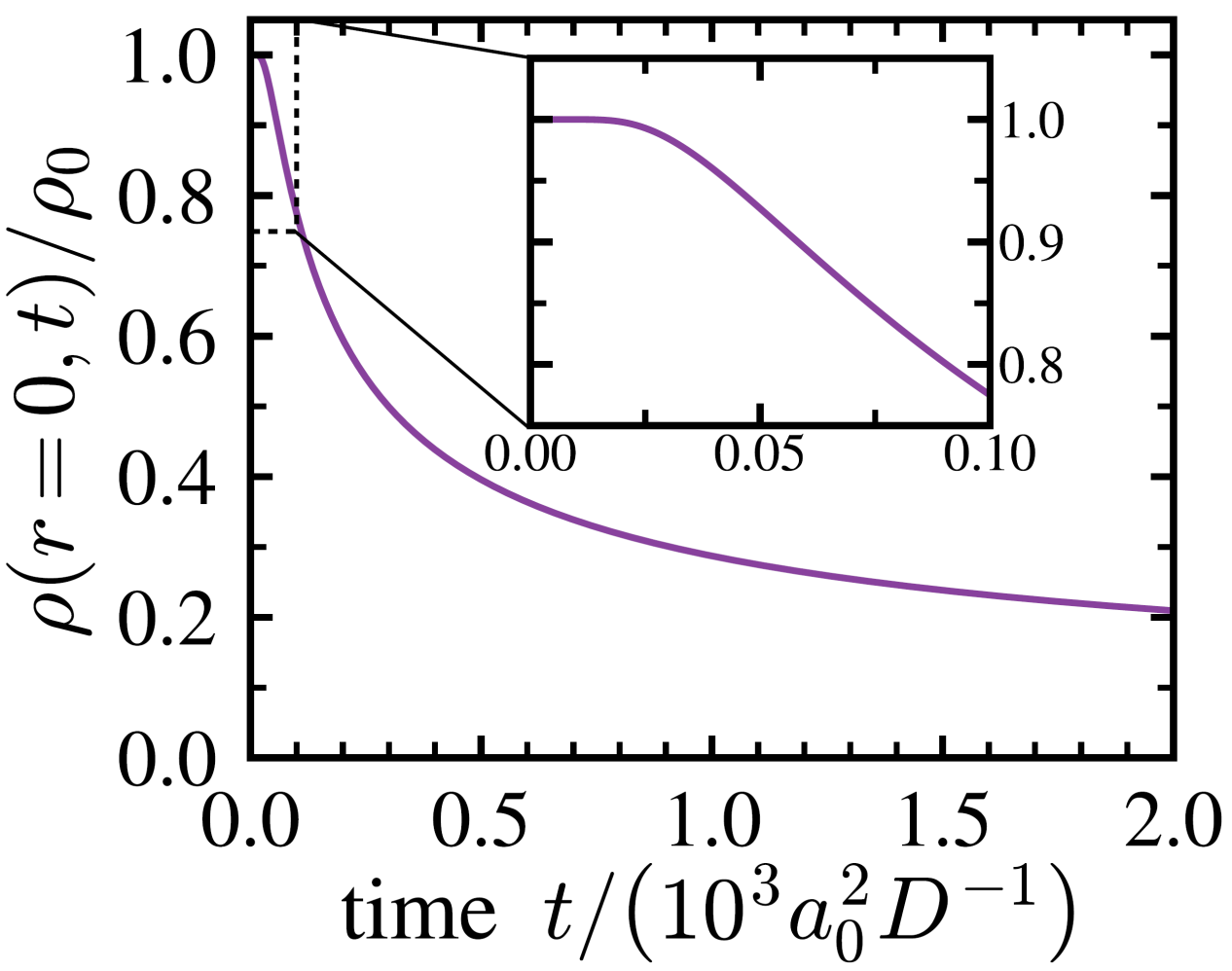

**Fig. S11 Evolution of the central density in the model calculations.** Shown is the result of model calculations using the above standard input parameters. The inset shows an enlargement of the initial core density evolution showing the minimally delayed onset of density reduction.

**References**

---